\documentclass[aps,prd,showpacs,nofootinbib,tightenlines]{revtex4}
\usepackage[colorlinks,urlcolor=blue,linkcolor=blue,citecolor=blue]{hyperref} %
\usepackage{amsmath,amssymb,epsfig,graphicx,float,color,booktabs,ulem,verbatim,times} %
\usepackage[utf8]{inputenc}
\begin{document}
\title{\Large{Contributions of the subprocess $K^*_0(1430) \to K\eta^{\prime}$
                      in the charmless three-body $B$ meson decays}\vspace{0.4cm}}

\author{Ai-Jun Ma$^{1}$}\email{theoma@163.com}
\author{Wen-Fei Wang$^{2,3}$} \email{wfwang@sxu.edu.cn}

\affiliation{\small
                     $^1$School of Mathematics and Physics, Nanjing Institute of Technology,
                            Nanjing, Jiangsu 211167, China \\
                    $^2$Institute of Theoretical Physics and State Key Laboratory of Quantum Optics and Quantum Optics Devices,
                            Shanxi University, Taiyuan, Shanxi 030006, China \\
                    $^3$Departament de F\'{\i}sica Qu\`antica i Astrof\'{\i}sica and Institut de Ci\`encies del Cosmos (ICCUB),
                            Facultat de F\'{\i}sica, Universitat de Barcelona, Mart\'i i Franqu\`es 1, 08028, Barcelona, Spain }

\date{\today}
%XXXXXXXXXXXXXXXXXXXXXXXXXXXXXXXXXX

\begin{abstract}
We study the contributions for $K\eta^{\prime}$ pair originating from the scalar intermediate state $K_0^{*}(1430)$
in the three-body decays $B\to K\eta^{\prime} h$ ($h=\pi, K$) within the perturbative QCD approach. The contribution
of $K^*_0(1430)\to K\eta^{\prime}$ is described by the Flatt${\rm \acute{e}}$ formula with coupled
channels $K\pi$, $K\eta$, and $K\eta^{\prime}$.  The strong coupling constants $g_{K^*_0K\eta^{(\prime)}}$
are extracted from $g_{K^*_0 K\pi}$ within flavor SU$(3)$ symmetry. In spite of the strong depression
by phase space near the threshold of $K\eta^\prime$, the $CP$ averaged branching fractions for the
$B\to K^*_0(1430) h \to K\eta^\prime h$ decays are predicted to be on the order of $10^{-8}$ to $10^{-5}$, which
are non-negligible for the corresponding three-body $B$ decays. Since the K$\eta$ system is almost decoupled from
the even-spin strange mesons under flavor SU$(3)$ symmetry,  those quasi-two-body $B$ decays with subprocess
$K^*_0(1430) \to K \eta$ shall have quite small branching ratios and are not taken into account in this work. We also
estimate that the branching fraction for $K_0^{*}(1430)\to K\eta^{\prime}$ is about one fifth of that for
$K_0^{*}(1430)\to K\pi$. The predictions for the relevant decays are expected to be tested by the LHCb and
Belle-II experiments in the future.
\end{abstract}


\maketitle

%XXXXXXXXXXXXXXXXXXXXXXXXXXXXXXXXXX% @ Begin

\section{Introduction}
It is known that the identification of the scalar resonances is experimentally difficult and the underlying inner structure
of the scalar mesons remains debated on the theoretical side. The light scalars, especially those with the mass below $1$ GeV,
are generally suggested as multiquark states, hadronic molecules, or glueballs, etc., besides the conventional quark-antiquark
configuration~\cite{jpg28-R249,pr389-61,pr397-257,pr454-1,pr658-1}. The resonance $K^*_0(1430)$, the least controversial one
in the light scalar states, has been observed decaying dominantly to the $K\pi$ pair for more than three decades~\cite{npb296-493}
and can be constructed as a $q\bar{q}$ state with $J^{PC} = 0^{++}$~\cite{PDG2022}.
In recent years, the other two coupled channels $K\eta$ and $K\eta^\prime$ for $K^*_0(1430)$ have also been measured by
different collaborations. For example, in Ref.~\cite{prd89-112004} the \textit{BABAR} collaboration reported the first observation
of the $K^*_0(1430) \to K\eta$ decay in the process $\eta_c \to \eta K^+K^-$ and gave a relative branching fraction ($\mathcal{B}$)
\begin{equation}
 R_{K\eta}=\frac{\mathcal{B}(K^*_0(1430) \to K\eta)}{\mathcal{B}(K^*_0(1430) \to K\pi)}=(9.2 \pm 2.5^{+1.0}_{-2.5})\%.
\end{equation}
Meanwhile, the $K^*_0(1430) \to K\eta^\prime$ decay was firstly observed in the measurement of
$\chi_{cJ} \to \eta^{\prime}K^+K^-$ by the BESIII collaboration~\cite{prd89-074030}.
Very recently, the Dalitz plot analysis for $\eta_c\to \eta^\prime K^+K^-$ was performed and the ratio
$R_{K\eta^\prime}=\frac{\mathcal{B}(K^*_0(1430) \to K\eta^\prime)}{\mathcal{B}(K^*_0(1430) \to K\pi)}$ was measured to
be quite large, with the value of $(39.7 \pm 6.4\pm 5.4)\%$ by the \textit{BABAR} collaboration~\cite{prd104-072002}.

Since the scalar states decay mainly into two pseudoscalars, three-body $B$ meson decays  provide us rich opportunities to
study the scalar resonant states. The contribution of $K^*_0(1430)$ in the $S$-wave $K\pi$ system has been solidly established in the measurements of the charmless three-body
$B$ meson decays by the Belle~\cite{prd71-092003,prd75-012006,prl96-251803}, \textit{BABAR}~\cite{prd72-072003,prd76-071103,prd78-012004,prd80-112001,prd83-112010,prd96-072001},
and LHCb collaborations~\cite{JHEP06-114,prl123-231802}, and
the corresponding branching fractions and direct $CP$ violations for the quasi-two-body channels $B\to K^*_0(1430)h \to K\pi h$ ($h=\pi, K$) have also been presented via amplitude analysis.
The $K^*_0(1430)$ contributions in most of the concerned three-body $B$ meson decays are found to be considerably large.
Particularly, the quasi-two-body components with $K^*_0(1430)$ in the decays $B^+ \to K^+\pi^+\pi^-$~\cite{prd71-092003,prl96-251803},
$B^0 \to K^0\pi^+\pi^-$~\cite{prd75-012006}, $B^0 \to K^+\pi^-\pi^0$~\cite{prd83-112010}, and $B^+ \to K^0\pi^+\pi^0$~\cite{prd96-072001}
were shown to be larger than $50\%$ of the corresponding total branching fractions. Theoretical researches were carried out in parallel.
As discussed in Refs.~\cite{plb561-258,npb899-247,JHEP10-117}, the factorization methods for quasi-two-body $B$ meson decays are still valid,
and a number of related works within different approaches~\cite{prd79-094005,prd81-094033,prd88-114014,prd89-074025,prd94-094015,
prd102-053006,prd103-036017,prd89-094007,prd99-076010,JHEP03-162,epjc80-517} based on factorization theorem have been done in order
to explain the data or provide theoretical predictions.
In addition, the resonant contributions including $K^*_0(1430)$ in $B^- \to K^+K^-\pi^-$ were considered by adopting the light-cone sum rule
approach~\cite{epjc82-113}, where the matrix element $\langle R, M\left | {\cal O}_i\right|B\rangle$ as a whole was calculated.
But, there are not many discussions on the contributions from subprocesses $K^*_0(1430) \to K\eta^{(\prime)}$ for the charmless three-body $B$ meson
decays on both the theoretical and experimental sides.

Within the flavor SU($3$) symmetry, the $K\eta$ branching fraction for the strange states like $K^*$ with even spin is expected
to be strongly suppressed, but for odd-spin states it will be quite substantial~\cite{prl46-1307,plb254-247,plb201-169}.
This is supported by the data ${\mathcal B}(K^*_2(1430) \to K\eta)=(0.15^{+0.34}_{-0.10})\%$ and ${\mathcal B}(K^*_3(1780) \to K\eta)=(30\pm13)\%$
presented by the {\it Particle Data Group} (PDG)~\cite{PDG2022}.
Since the $K\eta$ system almost decouples from the even-spin $K^*$ resonances, the relatively large $R_{K\eta}$ for $K^*_0(1430)$ presented by
\textit{BABAR}~\cite{prd89-112004} is not in complete agreement with the SU($3$) expectation and needs further confirmation.
In addition, there is no significant amplitude observed for the $K^*_0(1430)\to K\eta$ decay in the study of $K^-p \to K^-\eta p$ interaction
by LASS collaboration~\cite{plb201-169}.
The situation will be reversed for the $K\eta^{\prime}$ channel and the even-spin $K^*$ states are expected to couple preferentially to $K\eta^{\prime}$.
In the description of $K^*_0(1430)$ contribution with coupled channels, the $K\pi$ and $K\eta^{\prime}$ systems are included while the term for $K\eta$
is usually ignored in the literature~\cite{plb632-471,prd81-014002,prd78-052001,prd89-074030,prd104-072002,prd104-012016,JHEP08-196,plb653-1}.
In this context, we will focus on the contributions from $K^*_0(1430)$ for the $K\eta^{\prime}$ final state in the
three-body decays $B\to K\eta^{\prime} h$ in this work.

The perturbative QCD (PQCD) approach~\cite{prd63-054008,plb504-6,prd63-074009} based on the $k_T$ factorization has been widely adopted
in the investigations of the quasi-two-body $B$ meson decay processes in recent years~\cite{epjc80-517,JHEP03-162,prd91-042024,plb763-29,prd101-111901,
prd103-056021,epjc80-815,epjc77-199,prd97-033006,prd98-113003,epjc79-792,epjc80-394,epjc82-1076,prd95-054008,epjc79-37,prd101-016015,
npb923-54,epjc79-539,prd103-016002,prd103-033003,prd107-013001}.
Employing the same method, we have studied the whole sixteen quasi-two-body decays with the type of $B\to K^*_0(1430)h \to K\pi h$ in
Ref.~\cite{JHEP03-162} and found that the PQCD predictions for the relevant decays agree well with the existing experimental results
from the \textit{BaBar}, Belle, and LHCb collaborations.
Now  we will extend the previous work to the calculation of the contributions from the subprocess $K^*_0(1430) \to K\eta^{\prime}$
in the charmless three-body $B\to K\eta^{\prime} h$ decays.
In Ref.~\cite{prd104-072002}, evidence for $K^*_0(1950)$ in the $\eta_c\to \eta^\prime K^+K^-$ decay was also found and the width was
measured as $\Gamma_{K^*_0(1950)}=(80 \pm32 \pm20)$ MeV, which is quite different from the commonly used value $(201\pm34\pm79)$ MeV
presented by LASS~\cite{npb296-493}.
Due to the lack of the well-defined distribution amplitudes for $K^*_0(1950)$ and the still large uncertainty of its parameters,
we shall leave it to the future studies.

The rest of this paper is organized as follows. In Sec. II, we briefly describe the PQCD framework for the quasi-two-body
$B\to K^*_0(1430)h \to K\eta^{\prime} h$ decays. In Sec. III, we provide the numerical results and give necessary discussions.
The summary is presented in Sec. IV.

%%%===========================%%%
\section{Framework}
\begin{figure}[tbp]
\centerline{\epsfxsize=14cm \epsffile{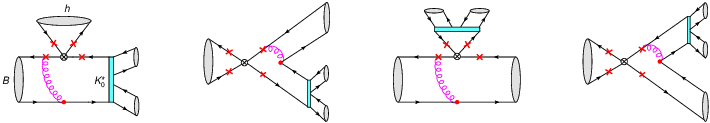}}
\caption{Typical Feynman diagrams for the quasi-two-body decay processes $B\to K_0^*(1430) h\to K\eta^\prime h$. The symbol $\otimes$ represents
         the insertion of the four-fermion vertices in the effective theory and $\times$ denotes the possible attachments of hard gluons.}
\label{fig-feyndiag}
\end{figure}
In the PQCD approach, the decay amplitude of the hadronic $B$ meson decay is expressed as the convolution of a hard kernel with the
distribution amplitudes (or wave functions) for the initial $B$ meson and the final state hadrons.
For a quasi-two-body decay, the two-meson distribution amplitudes are introduced to describe the interaction between the meson pair
which proceeds by the intermediate state.
Then, the factorization formula of the decay amplitude ${\mathcal A}$ for the $B\to K^*_0(1430)h \to K\eta^{\prime} h$ decay has the
form~\cite{plb561-258,plb763-29,JHEP03-162}
\begin{eqnarray}
{\mathcal A}=\phi_B \otimes {\mathcal H} \otimes  \phi^{S\text{-wave}}_{K\eta^{\prime}} \otimes \phi_{h},
\label{def-DA-Q2B}
\end{eqnarray}
where the hard kernel ${\mathcal H}$ contains only one hard gluon exchange.
The symbols $\phi_B$, $\phi^{S\text{-wave}}_{K\eta^{\prime}}$, and $\phi_h$ represent the distribution amplitudes for the $B$ meson, the $K\eta^{\prime}$
pair, and the bachelor meson $h$ ($\pi$ or $K$) respectively, which absorb the nonperturbative dynamics in the hadronization processes.
In this work, the same distribution amplitudes for the heavy $B$ meson, and the light $K$ and $\pi$ mesons are adopted as those listed in
Refs.~\cite{JHEP03-162,prd103-056021} and the references therein.

For the $K\eta^{\prime}$ system, the scalar form factor $F^{K\eta^{\prime}}_0(s)$ can be written via the following matrix element~\cite{npb622-279,prd101-034010}
\begin{eqnarray}
\langle K\eta^{\prime} |\bar{q}_2 q_1| 0 \rangle =C_{K\eta^{\prime}} \frac{\Delta_{K\pi}}{m_{q_2}-m_{q_1}} F_0^{K\eta^{\prime}}(s)= C_{K\eta^{\prime}}B_0 F_0^{K\eta^{\prime}}(s),
\label{eq-SKeta}
\end{eqnarray}
where the isospin factor $C_{K\eta^{\prime}}=\frac{2}{\sqrt{3}}$, the mass-squared difference $\Delta_{K\pi}=m_K^2-m_{\pi}^2$, the quarks $q_1=s$
and $q_2=(u,d)$ for $K=(K^+,K^0)$, and $q_1=(u,d)$ and $q_2=s$ for $K=(K^-,\bar{K^0})$.
When there is only the intermediate state $K^*_0(1430)$ for the $K\eta^{\prime}$ system, one also has~\cite{prd80-054007}
\begin{eqnarray}
\langle K\eta^{\prime} | \bar{q}_2 q_1| 0 \rangle =  \langle K\eta^{\prime} |  K_0^* \rangle \frac{1}{\mathcal{D}_{K^*_0}}
 \langle  K_0^* |\bar{q}_2 q_1 |0 \rangle = \Pi_{K_0^*K\eta^{\prime}}(s) \langle  K_0^{*} |\bar{q}_2 q_1  |0 \rangle,
\label{eq-SKeta1}
\end{eqnarray}
where $\langle K\eta^{\prime} |  K_0^* \rangle$ is the coupling constant for
the resonance $K_0^*(1430)$ with $K\eta^{\prime}$ and $\frac{1}{\mathcal{D}_{K^*_0}}$ stands for the propagator.
Then, we have the vertex function related form factor
\begin{eqnarray}
F_0^{K\eta^{\prime}}(s)=\frac{ \Pi_{K_0^*K\eta^{\prime}}(s) \langle  K_0^{*} |\bar q_2 q_1  |0 \rangle}{C_{K\eta^{\prime}} B_0}
=\frac{g_{K_0^*K\eta^{\prime}}m_{K_0^*}\bar f_{K_0^*}}{ C_{K\eta^{\prime}} B_0\mathcal{D}_{K^*_0}} \;,
\label{eq-SKeta2}
\end{eqnarray}
by using the definition of the scalar decay constant $\langle  K_0^{*} |\bar{q}_2 q_1|0 \rangle=m_{K_0^*}\bar f_{K_0^*}$.
The value of $\bar f_{K_0^*}$ is related to the vector decay constant $f_{K_0^*}$ through $\bar f_{K_0^*}= \frac{m_{K_0^*}f_{K_0^*}}{m_{q_2}(\mu)-m_{q_1}(\mu)}$ and
the result $f_{K_0^*(1430)}m^2_{K_0^*(1430)}=(0.0842\pm0.0045)$ GeV$^3$~\cite{plb462-14} is employed in this work.

The strong coupling constant $g_{K_0^*K\pi}$ can be determined from the measured partial width $\Gamma_{{K^*_0}\to K\pi}$
with the relation~\cite{prd88-114014,prd102-053006}
\begin{eqnarray}
\Gamma_{{K^*_0}\to h_1h_2}=\frac{q_0}{8\pi m^2_{K^*_0}}g^2_{K_0^*h_1h_2}\;.
\label{eq-gKpi}
\end{eqnarray}
The $q_0$ denotes the value at $s=m^2_{K_0^{*}}$ for the magnitude of the momentum for the daughter $h_1$ or $h_2$  which is defined as
\begin{eqnarray}
q=\frac{1}{2}\sqrt{\left[s-(m_{h_1}+m_{h_2})^2\right]\left[s-(m_{h_1}-m_{h_2})^2\right]/s}\;
\label{eq-q}
\end{eqnarray}
in the rest frame of the resonance $K_0^{*}(1430)$.
But in view of the insufficient studies on the partial width for $K_0^{*}(1430) \to K\eta^{\prime}$ and the fact that the pole mass of
$K_0^{*}(1430)$ is slightly smaller than the threshold for $K\eta^{\prime}$ in the ${\it Review~of~Particle~Physics}$~\cite{PDG2022},
it is not suitable to calculate the strong coupling constant for $K\eta^{\prime}$ with the above formulae directly.
We treat it under the flavor SU(3) symmetry and extract $g_{K^*_0K\eta^{\prime}}$, together with $g_{K^*_0K\eta}$, from the relations
\begin{eqnarray}
\label{eq-gKeta} \frac{g_{K^*_0K\eta}}{g_{K^*_0K\pi}}&=&\sqrt{\frac{1}{3}} \cos{\phi}-\sqrt{\frac{2}{3}} \sin{\phi}=-0.070, \\
\label{eq-gKetap}\frac{g_{K^*_0K\eta^\prime}}{g_{K^*_0K\pi}}&=&\sqrt{\frac{1}{3}} \sin{\phi}+\sqrt{\frac{2}{3}} \cos{\phi}=0.998,
\end{eqnarray}
with $g_{K^*_0K\pi}$ obtained from Eq.~(\ref{eq-gKpi}).
The mixing angle $\phi=39.3^{\circ}$~\cite{plb449-339,prd58-114006} in the quark flavor basis is employed while the physical
$\eta$ and $\eta^\prime$ are known as the mixtures from $\eta_q=\frac{u\bar{u}+d\bar{d}}{\sqrt{2}}$ and $\eta_s=s\bar{s}$ through the relation
\begin{equation}
\left(\begin{array}{c} \eta \\ \eta^{\prime} \end{array} \right)= \left(\begin{array}{cc}  \cos{\phi} & -\sin{\phi} \\ \sin{\phi} & \cos{\phi} \\ \end{array} \right)
 \left(\begin{array}{c} \eta_q \\ \eta_s \end{array} \right).
\end{equation}
According to the discussions in Refs.~\cite{plb632-471,prd81-014002,prd78-052001,prd89-074030,prd104-072002,prd104-012016,JHEP08-196}, we adopt the
Flatt${\rm \acute{e}}$ parametrization~\cite{plb63-224} for the denominator of the propagator
\begin{eqnarray}
\label{flatte} {\mathcal{D}_{K^*_0}}=m^2_{K^*_0}-s-i  [g^2_{K\pi} \rho_{K\pi}(s)+g^2_{K\eta} \rho_{K\eta}(s)+g^2_{K\eta^\prime} \rho_{K\eta^\prime}(s)],
\end{eqnarray}
where the phase space factor $\rho_{ab}(s)=\frac{2q}{\sqrt{s}}$. It should be noted that the definition of coupling constants $g_{ab}$ here
is different by a factor of $\frac{1}{4\sqrt{\pi}}$ from the $g_{K^*_0ab}$ shown in Eqs.~(\ref{eq-gKeta}) and (\ref{eq-gKetap}).
One can find different values of $g^2_{K\pi}$, $g^2_{K\eta}$, and $g^2_{K\eta^\prime}$ in Refs.~\cite{plb632-471,prd78-052001,prd104-072002,
prd81-014002,epja59-79}, which are listed in Table~\ref{g2}.
We employ $g^2_{K\pi}=(0.412\pm0.130)~{\rm GeV^2}$, $g^2_{K\eta}=(0.00204\pm0.00064^{+0.00113}_{-0.00089})~{\rm GeV^2}$, and $g^2_{K\eta^\prime}=(0.410\pm0.129\pm0.001)~{\rm GeV^2}$ in this work through the relation
\begin{equation}
   g^2_{K\pi(K\eta^{(\prime)})}=\frac{g^2_{K^*_0K\pi(K\eta^{(\prime)})}}{16\pi},
\end{equation}
where the first error comes from the uncertainty of $\Gamma_{K_0^{*}(1430)\to K\pi}$ with
$\Gamma_{K_0^{*}(1430)}=(270\pm80)~{\rm MeV}$ and $\mathcal{B}(K_0^{*}(1430)\to K\pi)=(93\pm10)\%$~\cite{PDG2022},
and the second error for $g^2_{K\eta}$ and $g^2_{K\eta^{\prime}}$ arises from the mixing angle $\phi=(39.3\pm1.0)^{\circ}$~\cite{plb449-339,prd58-114006}.

\begin{table}[thb]   %%[H] %%[thb]
\begin{center}
\caption{Comparison of parameters $g^2_{K\pi}$, $g^2_{K\eta}$, and $g^2_{K\eta^\prime}$ fitted in different literature.}
\label{g2}
\begin{tabular}{c c c c } \hline\hline
    ~$g^2_{K\pi}~({\rm GeV^2})$~~ & ~~$g^2_{K\eta}~({\rm GeV^2})$~~ & ~~$g^2_{K\eta^\prime}~({\rm GeV^2})$~~ & ~Refs~   \\ \hline  %
    ~$0.458\pm0.032\pm0.044$~~&~~0~~&~~$(1.50\pm0.24\pm0.24)g^2_{K\pi}$~~&~\cite{prd104-072002}~\\
    ~0.353~~&~~0~~&~~1.15$g^2_{K\pi}$~~& ~\cite{plb632-471}~\\
    ~$0.284\pm0.009~(0.299\pm0.005)~~$~~&~~0~~&~~$0.039\pm0.053~(0.053\pm0.016)$~~&~\cite{prd78-052001}~\\
    ~$0.284\pm0.012~~$~~&~~0~~&$(0.62\pm0.06)g^2_{K\pi}$~~&~\cite{prd81-014002}~\\
    ~0.515~~&~~0.030~~&~~0.671~~&~\cite{epja59-79}~\\
    ~$0.412\pm0.130~~$~~&~~$0.00204\pm0.00064^{+0.00113}_{-0.00089}$~~&~~$0.410\pm0.129\pm0.001$~~&~This work~\\ \hline\hline
\end{tabular}
\end{center}
\end{table}

The $S$-wave $K\eta^{\prime}$ distribution amplitudes along with the subprocess $K^*_0(1430) \to K\eta^{\prime}$ are defined in
the same way as those for the $K\pi$ system~\cite{JHEP03-162}
\begin{eqnarray}
\Phi_{K\eta^{\prime}}(z,s)=\frac{1}{\sqrt{2N_c}}\left[{p\hspace{-1.6truemm}/}\phi(z,s)
+\sqrt s\phi^{s}(z,s)+\sqrt s({v \hspace{-1.7truemm}/ }{n \hspace{-1.8truemm}/}-1)
\phi^{t}(z,s) \right],
\label{def-wavefun-Kpi}
\end{eqnarray}
with the momentum $p$ for the $K\eta^{\prime}$ pair, the momentum fraction $z$ for the spectator quark, the squared invariant mass
$s=p^2=m^2_{K\eta^{\prime}}$ and the dimensionless vectors $v=(0,1,{0}_{\rm T})$ and $n=(1,0,{0}_{\rm T})$.
The twist-$2$ and twist-$3$ light-cone distribution amplitudes are parametrized as~\cite{plb730-336,prd73-014017,prd87-114001}
\begin{eqnarray}
\phi(z,s)&=&\frac{F_{K\eta^{\prime}}(s) }{2\sqrt{2N_c}}\left\{ 6z(1-z)\left[a_{0}(\mu)+
\sum^{\infty}_{m=1} a_{m}(\mu)C^{3/2}_m(2z-1) \right]\right\}\;, \\
\phi^{s}(z,s)&=&\frac{F_{K\eta^{\prime}}(s) }{2\sqrt{2N_c}}\;,\quad\quad \\
\phi^{t}(z,s)&=&\frac{F_{K\eta^{\prime}}(s) }{2\sqrt{2N_c}}(1-2z)\;,
\label{def-wavefun-twist2}
\end{eqnarray}
where the factor $F_{K\eta^{\prime}}(s)$ is related to the scalar form factor $F^{K\eta^{\prime}}_0(s)$ by the equation
$F_{K\eta^{\prime}}(s)=\frac{C_{K\eta^{\prime}}B_0}{m_{K_0^*}} F_0^{K\eta^{\prime}}(s)$.
The symbols $C^{3/2}_m$ are the Gegenbauer polynomials and the value of $a_0$ equals to $(m_{(u,d)}(\mu)-m_s(\mu))/\sqrt s$ for ($K^{*+}_0, K^{*0}_0$)
and $(m_s(\mu)-m_{(u,d)}(\mu))/\sqrt s$ for ($K^{*-}_0, \bar K^{*0}_0$).
For the Gegenbauer moments $a_m$, the odd terms with $a_1=-0.57\pm0.13$ and $a_3=-0.42\pm0.22$ at the scale $\mu=1$ GeV are adopted and the even
terms could be neglected according to~\cite{prd73-014017}.

The differential branching fraction for the quasi-two-body decays $B\to K^*_0(1430)h \to K\eta^{\prime} h$ can be written as~\cite{PDG2022}
\begin{eqnarray}
\frac{d\mathcal{B}}{d\zeta}=\tau_B \frac{q_h q }
 {64 \pi^3 m_B }{|{\cal A}|^2} \label{eqn-diff-bra}
\end{eqnarray}
with the variable $\zeta=\frac{s}{m^2_B}$ and the $B$ meson mean lifetime $\tau_B$.
The magnitude of the momentum for the third meson $h$ besides the $K\eta^{\prime}$ pair is expressed as
\begin{eqnarray}
q_h=\frac{1}{2}\sqrt{\big[\left(m^2_{B}-m_{h}^2\right)^2 -2\left(m^2_{B}+m_{h}^2\right)s+s^2\big]/s}
\label{def-qh}
\end{eqnarray}
in the center-of-mass frame of the resonance $K^*_0(1430)$, where $m_h$ is the mass of the bachelor kaon or pion.
The direct $CP$ asymmetry ${\mathcal A}_{CP}$ is defined as
\begin{eqnarray}
{\mathcal A}_{CP}=\frac{{\mathcal B}(\bar B\to \bar f)-{\mathcal B}(B\to f)}{{\mathcal B}(\bar B\to \bar f)+{\mathcal B}(B\to f)}.
\end{eqnarray}
In addition, the decay amplitudes ${\cal A}$ in this work have the same expressions as the cases for quasi-two-body decays $B\to K^*_0(1430)h \to K\pi h$
except for the replacement $F_{K\pi}$ to $F_{K\eta^{\prime}}$ in the two-meson distribution amplitudes.
The explicit expressions of the decay amplitudes ${\cal A}$, together with the individual amplitudes for the factorizable and nonfactorizable
Feynman diagrams as shown in Fig.~\ref{fig-feyndiag} can be found in the Appendix of Ref.~\cite{JHEP03-162}.

%%%===========================%%%
\section{Results and discussions}

In the numerical calculations, the input parameters such as the QCD scale, masses, and decay constants, are summarized as follows
(in units of {\rm GeV})~\cite{PDG2022,prd76-074018}:
\begin{eqnarray}
\Lambda_{QCD}^{f=4}&=&0.250, \quad m_{B^\pm}=5.279,\quad m_{B^0}=5.280,\quad m_{B_s^0}=5.367,\nonumber\\
m_{\pi^\pm}&=&0.140,\quad m_{\pi^0}=0.135,~\quad m_{K^\pm}=0.494, \quad m_{K^0}=0.498,\nonumber\\
\quad m_{\eta}&=&0.548,\quad m_{\eta^\prime}=0.958,\quad~ m_{K_0^{*}(1430)}=1.425\pm 0.050,\nonumber\\
\quad f_{B}&=&0.190,\quad~ f_{B_s}=0.230,\quad~ f_{\pi}=0.130,\quad~ f_{K}=0.156. \label{eq:inputs}
\end{eqnarray}
The lifetimes of $B$ mesons are adopted as $\tau_{B^\pm} = 1.638 \times 10^{-12} \, {\rm s}$, $\tau_{B^0} = 1.519 \times 10^{-12} \, {\rm s}$,
and $\tau_{B_s} = 1.520 \times 10^{-12} \, {\rm s}$~\cite{PDG2022}. For the Wolfenstein parameters of the CKM matrix elements, we use the values
$A= 0.826^{+0.018}_{-0.015},~\lambda= 0.22500\pm0.00067$, $\bar{\rho} = 0.159\pm0.010$, and $\bar{\eta}= 0.348\pm0.010$ provided by PDG~\cite{PDG2022}.

%%%%%%%%%%%%%-table-K(1430)-%%
\begin{table}[thb]   %%[H] %%[thb]
\begin{center}
\caption{PQCD predictions of the $CP$ averaged branching fractions and the direct $CP$
         asymmetries for the quasi-two-body $B\to K_0^*(1430)h \to K\eta^\prime h$ decays.}
\label{Res-1430Ketap}
\begin{tabular}{l c c } \hline\hline
 ~~~~~~~~~~~~~Decay modes     &    $\mathcal B$    &  ${\mathcal A}_{CP}$            \\
\hline  %
  $B^+\to  K_0^{*}(1430)^0 \pi^+\to K^0\eta^\prime \pi^+$    &~~$5.27^{+2.22+0.54+0.80+0.65}_{-1.43-0.28-0.75-0.87}\times 10^{-6}$~~~
      & ~~$-0.02^{+0.00+0.00+0.00+0.00}_{-0.00-0.01-0.00-0.00}$~~    \\
  $B^+\to K_0^{*}(1430)^+ \pi^0\to K^+\eta^\prime \pi^0$     &~~$2.01^{+0.77+0.23+0.34+0.23}_{-0.50-0.17-0.32-0.31}\times 10^{-6}$~~~
      & ~~$0.01^{+0.00+0.01+0.00+0.00}_{-0.01-0.01-0.00-0.00}$~~     \\
  $B^+\to K_0^{*}(1430)^+ \bar K^0\to K^+\eta^\prime \bar K^0$    &~~$1.03^{+0.13+0.77+0.17+0.12}_{-0.06-0.40-0.15-0.15}\times 10^{-7}$~~~
      & ~~$-0.21^{+0.02+0.04+0.06+0.02}_{-0.02-0.02-0.06-0.01}$~~    \\
  $B^+\to \bar K_0^{*}(1430)^0 K^+\to \bar K^0\eta^\prime K^+$    &~~$8.41^{+2.22+2.59+1.29+1.00}_{-1.46-2.16-1.19-1.34}\times 10^{-7}$~~~
      & ~~$0.35^{+0.06+0.07+0.03+0.01}_{-0.06-0.09-0.02-0.01}$~~    \\
\hline %
  $B^0\to K_0^{*}(1430)^+ \pi^-\to K^+\eta^\prime \pi^-$     &~~~$4.98^{+2.04+0.35+0.73+0.56}_{-1.31-0.44-0.69-0.79}\times 10^{-6}$~~~
      & ~~$0.01^{+0.01+0.01+0.00+0.00}_{-0.01-0.02-0.00-0.00}$~~    \\
  $B^0\to  K_0^{*}(1430)^0 \pi^0\to K^0\eta^\prime \pi^0$    &~~~$3.18^{+1.37+0.35+0.41+0.39}_{-0.89-0.14-0.40-0.53}\times 10^{-6}$~~~
      & ~~$-0.02^{+0.00+0.00+0.00+0.00}_{-0.00-0.01-0.00-0.00}$~~    \\
  $B^0\to K_0^{*}(1430)^+  K^-\to K^+\eta^\prime K^-$        &~~~$1.30^{+0.12+1.06+0.11+0.07}_{-0.09-0.56-0.02-0.13}\times 10^{-8}$~~~
      & ~~$-0.03^{+0.03+0.26+0.22+0.01}_{-0.00-0.21-0.29-0.01}$~~    \\
  $B^0\to K_0^{*}(1430)^- K^+\to K^-\eta^\prime K^+$         &~~~$1.62^{+0.24+1.07+0.06+0.11}_{-0.20-0.71-0.04-0.16}\times 10^{-7}$~~~
      & ~~$-0.11^{+0.02+0.08+0.01+0.01}_{-0.00-0.05-0.00-0.00}$~~    \\
  $B^0\to  K_0^{*}(1430)^0 \bar K^0\to K^0\eta^\prime \bar K^0$    &~~~$1.33^{+0.12+1.02+0.18+0.13}_{-0.06-0.62-0.14-0.16}\times 10^{-7}$~~~
      & ~~ - ~~    \\
  $B^0\to \bar K_0^{*}(1430)^0 K^0\to \bar K^0\eta^\prime K^0$     &~~~$8.32^{+2.15+2.94+1.19+0.96}_{-1.49-2.42-1.11-1.27}\times 10^{-7}$~~~
      & ~~ - ~~    \\
\hline %
  $B_s^0\to  K_0^{*}(1430)^-  \pi^+\to K^-\eta^\prime \pi^+$       &~~~$1.04^{+0.40+0.18+0.00+0.09}_{-0.28-0.17-0.00-0.13}\times 10^{-5}$~~~
      & ~~$0.18^{+0.03+0.04+0.01+0.00}_{-0.03-0.05-0.01-0.00}$~~    \\
  $B_s^0\to \bar K_0^{*}(1430)^0 \pi^0\to \bar K^0\eta^\prime \pi^0$     &~~~$1.17^{+0.41+0.58+0.10+0.10}_{-0.26-0.43-0.09-0.15}\times 10^{-7}$~~~
      & ~~$0.78^{+0.04+0.00+0.02+0.00}_{-0.07-0.05-0.02-0.00}$~~    \\
  $B_s^0\to  K_0^{*}(1430)^+  K^-\to K^+\eta^\prime  K^-$          &~~~$5.62^{+1.23+3.02+0.83+0.58}_{-0.65-2.23-0.76-0.79}\times 10^{-6}$~~~
      & ~~$0.04^{+0.01+0.01+0.01+0.00}_{-0.01-0.02-0.01-0.00}$~~    \\
  $B_s^0\to K_0^{*}(1430)^- K^+\to K^-\eta^\prime K^+$             &~~~$6.17^{+1.08+4.33+0.90+0.46}_{-0.73-3.14-0.80-0.71}\times 10^{-6}$~~~
      & ~~$-0.45^{+0.02+0.09+0.04+0.00}_{-0.02-0.14-0.04-0.00}$~~    \\
  $B_s^0\to  K_0^{*}(1430)^0 \bar K^0\to K^0\eta^\prime \bar K^0$  &~~~$5.52^{+1.24+2.91+0.86+0.62}_{-0.68-1.53-0.78-0.83}\times 10^{-6}$~~~
      & ~~ - ~~    \\
  $B_s^0\to \bar K_0^{*}(1430)^0 K^0\to \bar K^0\eta^\prime K^0$   &~~~$4.71^{+0.63+3.71+0.78+0.36}_{-0.43-2.56-0.69-0.55}\times 10^{-6}$~~~
      & ~~ - ~~    \\
\hline\hline
\end{tabular}
\end{center}
\end{table}
%%%%%%%%%%%%%-table-K(1430)-%%

Within the framework of PQCD approach, we calculate the $CP$ averaged branching fractions and the direct $CP$ asymmetries
for the concerned quasi-two-body decays $B\to K^*_0(1430) h$ with the subprocess $K_0^*(1430) \to K\eta^\prime$, and list them in Table~\ref{Res-1430Ketap}.
For those PQCD predictions, the first error comes from the uncertainty of shape parameter $\omega_B=0.40\pm0.04$ or $\omega_{B_s}=0.50\pm0.05$
in the $B^{+,0}$ or $B_s^0$ meson distribution amplitudes;
the second error is induced by the uncertainties of the Gegenbauer moments $a_1=-0.57\pm0.13$ and $a_3=-0.42\pm0.22$ in the distribution
amplitudes for the $K\eta^\prime$ system;
the third one is due to the chiral masses $m^K_0=(1.6\pm0.1)~{\rm GeV}$ and $m^{\pi}_0=(1.4\pm0.1)~{\rm GeV}$, and the Gegenbauer moment
$a^h_2 = 0.25\pm0.15$ of the bachelor pion or kaon; and the last one is caused by $\Gamma_{K_0^{*}(1430)}=(270\pm80)~{\rm MeV}$ and
$\mathcal{B}(K_0^{*}(1430)\to K\pi)=(93\pm10)\%$. The errors coming from the uncertainties of other parameters are small and have been neglected.
One interesting thing is that the large uncertainties of the decay width $\Gamma_{K_0^{*}(1430)}$ and $\mathcal{B}(K_0^{*}(1430)\to K\pi)$
result in an error around $10\%$ for the branching fractions of the considered quasi-two-body decays.
The reason is that the width $\Gamma_{K_0^{*}(1430)}$ and $\mathcal{B}(K_0^{*}(1430)\to K\pi)$ are used to calculate the coupling constant
$g_{K_0^{*}K\pi}$ which appears in both the numerator and the denominator of the form factor $F_0^{K\eta^{\prime}}(s)$; the effects of the
variation of $\Gamma_{K_0^{*}(1430)}$ and $\mathcal{B}(K_0^{*}(1430)\to K\pi)$ on the branching fractions are partially canceled out.

%%%%%%%%%%%%%-table-phi-dependence-%%
\begin{table}[thb]   %%[H] %%[thb]
\begin{center}
\caption{The $\phi$ dependence of the PQCD-predicted branching ratio and direct $CP$ asymmetry for the $B^+\to K_0^{*}(1430)^0 \pi^+\to K^0\eta^\prime \pi^+$ decay. }
\label{phi-dependence}   %% ^{+}_{-}
%\small %\footnotesize
\begin{tabular}{c c c c c |c} \hline\hline
   $\phi$  & $38^{\circ}$ &   $40^{\circ}$ &$42^{\circ}$ & $44^{\circ}$& $39.3^{\circ}$ \\
\hline  %
  $\mathcal B(10^{-6})$    &~~~5.29~~~&~~~5.26~~~&~~~5.22~~~&~~~5.16~~~&~~~~~5.27~~~~~ \\
  ${\mathcal A}_{CP}$     &~~~-0.02~~~&~~~-0.02~~~&~~~-0.02~~~&~~~-0.02~~~&~~~~~-0.02~~~~~ \\
\hline\hline
\end{tabular}
\end{center}
\end{table}
%%%%%%%%%%%%%-table-phi-dependence-%%

For the $\eta-\eta^{\prime}$ mixing angle, we follow the prediction of $\phi=(39.3\pm1.0)^{\circ}$ from Refs.~\cite{plb449-339,prd58-114006} in this work.
Actually, the value of $\phi$ has also been investigated in many phenomenological and experimental works
~\cite{epjc7-271,prd60-114012,plb503-271,prd71-014034,jhep06-029,epjc50-771,jhep05-006,plb648-267,jhep07-105,jhep01-024,prd108-092003,2307.12852},
with results ranging from $38^{\circ}$ to $44^{\circ}$ approximately. Since the extraction of coupling constants $g_{K\eta^{(\prime)}}$ in this work is related to the
mixing angle $\phi$, it is necessary to check the $\phi$ dependence of the PQCD-predicted branching ratios and $CP$-violating asymmetries for the considered
quasi-two-body decays.
Taking the decay $B^+\to K_0^{*}(1430)^0 \pi^+\to K^0\eta^\prime \pi^+$ as an example, we calculate its branching ratio and direct $CP$ asymmetry
with four fixed values of $\phi$ in the range of $38^{\circ}$ to $44^{\circ}$, and list them in Table~\ref{phi-dependence}.
One can find that the PQCD predictions of these two physical observables $\mathcal B$ and ${\mathcal A}_{CP}$ are not very sensitive on the variation of the given value of $\phi$,
and this is because the change of input $\phi$ in the specified range has a weak effect on the crucial parameter $g_{K^*_0K\eta^{\prime}}$ extracted from the relation of Eq.~(\ref{eq-gKetap}).
The case of $K\eta$ channel in Eq.~(\ref{eq-gKeta}) is quite different, and a considerable error for $g^2_{K\eta}$ caused by $\phi=(39.3\pm1.0)^{\circ}$
can be found in the last line of Table~\ref{g2}.
When we fix $\phi$ as $42^{\circ}$ and $44^{\circ}$, the values of $g^2_{K\eta}$ are calculated to be $0.00567~{\rm GeV^2}$ and $0.00951~{\rm GeV^2}$, respectively, which are almost three and five times
the central value of $0.00204~{\rm GeV^2}$ obtained by employing $\phi=39.3^{\circ}$.
In this respect, more precise theoretical and experimental studies on the coupling constant for $K_0^{*}(1430) \to K\eta$ are needed in the future, and it would also be helpful
to the determination of mixing angle $\phi$.
However, the variation of $g^2_{K\eta}$ in the Flatt${\rm \acute{e}}$ formula of Eq.~(\ref{flatte}) is negligible compared to the values of $g^2_{K\pi}$ and $g^2_{K\eta^{\prime}}$,
and does not have much effect on our predictions of the branching ratios and $CP$-violating asymmetries for the considered quasi-two-body decays.
Therefore, the small errors caused by the mixing angle $\phi$ used in this work are not taken into account in Table~\ref{Res-1430Ketap}.

In the charmless nonleptonic $B$ meson decays, the direct $CP$ violations arise from the interference between the decay amplitudes for
the tree and penguin diagrams.
As shown in Table~\ref{Res-1430Ketap}, there are no direct $CP$ asymmetries for the quasi-two-body decays $B_{(s)}^0\to \bar K_0^{*}(1430)^0 K^0\to \bar K^0\eta^{\prime} K^0$
and $B_{(s)}^0\to K_0^{*}(1430)^0 \bar K^0\to K^0\eta^{\prime} \bar K^0$ since these decays occur only through the penguin diagrams,
while the decays via the $b \to dq\bar{q}$ transition at the quark level generally have sizable direct $CP$ violations due to
the effect of parameters in the CKM matrix elements.
In this work, the $CP$ averaged branching fractions of the $B\to K^*_0(1430) h \to K\eta^\prime h$ decays are predicted to be on the order of $10^{-8}$
to $10^{-5}$ which are non-negligible as expected because of the large coupling constant for $K_0^{*}(1430) \to K\eta^{\prime}$.
Up to now, there are still not any experimental measurements or theoretical works have been presented for the relevant three-body or quasi-two-body
$B$ meson decays, and all of these predictions are expected to be tested by the future experiments from LHCb and Belle-II.

%%%%%%%%%%%%%-table-K(1430)-%%
\begin{table}[thb]   %%[H] %%[thb]
\begin{center}
\caption{The available experimental measurements for the branching fractions of the $B \to K_0^{*}(1430) h$ decays from ${\it Review~of~Particle~Physics}$. }
\label{2body-1430}   %% ^{+}_{-}
%\small %\footnotesize
\begin{tabular}{l c c} \hline\hline
   ~~~~~~Decay modes~~~~~~  & ~~~~~~Unit~~~~~~ &   ~~~~~~Data~\cite{PDG2022}~~~~~~   \\
\hline  %
  $B^+\to  K_0^{*}(1430)^0 \pi^+$    &~~~~~~$(10^{-5})$~~~~~~ & $3.9^{+0.6}_{-0.5}$    \\
  $B^+\to K_0^{*}(1430)^+ \pi^0$     &~~~~~~$(10^{-5})$~~~~~~ & $1.19^{+0.20}_{-0.23}$   \\
  $B^+\to \bar{K}_0^{*}(1430)^0 K^+$ &~~~~~~$(10^{-7})$~~~~~~ & $3.8\pm 1.3$ \\
  $B^0\to  K_0^{*}(1430)^+ \pi^-$    &~~~~~~$(10^{-5})$~~~~~~ & $3.3\pm 0.7$    \\
\hline\hline
\end{tabular}
\end{center}
\end{table}
%%%%%%%%%%%%%-table-K(1430)-%%

Within the quasi-two-body approximation, the branching ratios of the two-body decay $B\to K^*_0(1430) h$ and the related quasi-two-body decay
with cascade decay $K^*_0(1430) \to K h^{\prime}$ satisfy the factorization relation
\begin{eqnarray}\label{eq-approx}
 \mathcal {B}(B\to K^*_0(1430) h \to K h^{\prime} h)~\approx~\mathcal {B}(B\to K^*_0(1430) h ) \times \mathcal {B}(K^*_0(1430) \to K h^{\prime}).
\end{eqnarray}
In Table~\ref{2body-1430}, we list the available experimental data for the branching fractions of the $B \to K_0^{*}(1430) h$ decays in the
${\it Review~of~Particle~Physics}$~\cite{PDG2022} averaged from the measured results by Belle~\cite{prd75-012006,prl96-251803}, \textit{BABAR}
~\cite{prd78-012004,prd80-112001,prd96-072001}, and LHCb~\cite{prl123-231802}.
Considering the fact that the data of those two-body decays are extracted from the measured branching fractions for the related quasi-two-body
decays and $\mathcal{B}(K_0^{*}(1430)\to K\pi)=(93\pm10)\%$, together with the PQCD predictions
in this work, we obtain that the ratio for the central values of the branching fractions are
\begin{eqnarray}
\label{eq-ratio1} \frac{\mathcal {B}(B^+\to  K_0^{*}(1430)^0 \pi^+\to K^0\eta^{\prime} \pi^+)}{\mathcal {B}(B^+\to  K_0^{*}(1430)^0 \pi^+ \to  K\pi \pi^+)}=14.5\%, \\
\label{eq-ratio2} \frac{\mathcal {B}(B^+\to K_0^{*}(1430)^+ \pi^0\to K^+\eta^{\prime}\pi^0)}{\mathcal {B}(B^+\to K_0^{*}(1430)^+ \pi^0 \to K\pi \pi^0)}=18.2\%,  \\
\label{eq-ratio3} \frac{\mathcal {B}(B^0\to  K_0^{*}(1430)^+ \pi^- \to K^+\eta^{\prime}\pi^-)}{\mathcal {B}(B^0\to  K_0^{*}(1430)^+ \pi^- \to K\pi\pi^-)}=16.2\%.
\end{eqnarray}
For $\mathcal {B}(B^+\to \bar{K}_0^{*}(1430)^0 K^+)$, the experimental value is approximately one order of magnitude less than the theoretical
predictions in Refs.~\cite{prd91-074022,CTP53-540,JHEP03-162,prd105-016002} which also cannot be understood in this work.
By adopting our previous result $\mathcal {B}(B^+\to \bar{K}_0^{*}(1430)^0 K^+ \to K^-\pi^+ K^+)=(2.86\pm0.85) \times 10^{-6}$\cite{JHEP03-162}
and the relation $\mathcal {B}(\bar{K}_0^{*}(1430)^0 \to K^-\pi^+)=\frac{2}{3}\mathcal {B}(\bar{K}_0^{*}(1430)^0 \to K\pi)$, the same ratio becomes
\begin{eqnarray}\label{eq-ratio4}
\frac{\mathcal {B}(B^+\to \bar{K}_0^{*}(1430)^0 K^+ \to \bar{K}^0\eta^{\prime} K^+)}{\mathcal {B}(B^+\to \bar{K}_0^{*}(1430)^0 K^+ \to K\pi K^+)}=19.6\%.
\end{eqnarray}
Therefore, we estimate the value of $R_{K\eta^\prime}=\frac{\mathcal{B}(K^*_0(1430) \to K\eta^\prime)}{\mathcal{B}(K^*_0(1430) \to K\pi)}$ to be
close to $20\%$, which is about half of \textit{BABAR}'s result $(39.7 \pm 6.4\pm 5.4)\%$~\cite{prd104-072002} by using the ratio of measured
$\mathcal {B}(\eta_c  \to K^-K_0^{*}(1430)^+ \to K^-K^+\pi^0)$~\cite{prd89-112004} and $\mathcal {B}(\eta_c \to K^-K_0^{*}(1430)^+ \to K^-K^+\eta^{\prime})$~\cite{prd104-072002}
in the corresponding three-body decays, and the relation $ \mathcal {B}(K_0^{*}(1430)^+ \to K^+\pi^0)=\frac{1}{3} \mathcal {B}(K_0^{*}(1430)^+ \to K\pi)$.
Here, the prediction of $R_{K\eta^\prime}$ still has a large margin of error due to the considerable uncertainties for the branching fractions of
corresponding quasi-two-body decays by PQCD and experiments.
Besides, the validity of the factorization relation in Eq.~(\ref{eq-approx}) will also influence our results.
In Refs.~\cite{prd103-036017,plb813-136058}, the finite-width effects in three-body $B$ meson decays were discussed in detail and
a correction parameter $\eta_R$ was defined by
\begin{eqnarray}
\eta_R=\frac{\mathcal {B}(B \to Rh_3\to h_1h_2h_3)_{\Gamma_R \to 0}}{\mathcal {B}(B \to Rh_3\to h_1h_2h_3)}
      =\frac{\mathcal {B}(B \to Rh_3)\times\mathcal {B}(R \to h_1h_2)}{\mathcal {B}(B \to Rh_3\to h_1h_2h_3)}
      =1+\delta
\end{eqnarray}
where the correction $\delta$ was expected to be of order $\Gamma_R/{m_R}$.
Since the resonance $K^*_0(1430)$ has a large width, the finite-width effects should be taken into account for the corresponding decays.
Numerically, the parameter $\eta_R$ for $K^*_0(1430)$ within the framework of QCD factorization and the experimental parametrization for
the related three-body decay amplitudes were calculated to be $0.83\pm0.04$ and $1.11\pm0.03$, respectively.
It indicates that the two-body experimental results for $K^*_0(1430)$ which obtained in the narrow width approximation should be corrected by including the parameter $\eta_R$.
Meanwhile, the ratio $R_{K\eta^\prime}$ satisfies the relation
\begin{eqnarray}
R_{K\eta^\prime}=\frac{\mathcal{B}(K^*_0(1430) \to K\eta^\prime)}{\mathcal{B}(K^*_0(1430) \to K\pi)}=
\frac{\eta_{K^*_0K\eta^{\prime}}\times\mathcal {B}(B\to  K_0^{*}(1430) h \to K\eta^{\prime}h)}{\eta_{K^*_0K\pi}\times\mathcal {B}(B\to  K_0^{*}(1430) h \to K\pi h)}.
\end{eqnarray}
If we assume that the independent corrections $\delta$ for $\eta_{K^*_0K\eta^{\prime}}$ by PQCD and $\eta_{K^*_0K \pi}$ by experiments are both in the range of $-0.18$ to $0.18$
based on the $\eta^{\rm QCDF(EXPP)}_{K^*_0}$ in Refs.~\cite{prd103-036017,plb813-136058} and the value of $\Gamma_{K_0^{*}}/m_{K_0^{*}}$,
a maximum error of around $40\%$ will be introduced to the estimation of $R_{K\eta^\prime}$ which can be extracted by the ratio in Eqs.~(\ref{eq-ratio1})$\sim$(\ref{eq-ratio3}) and the factorization relation
in Eq.~(\ref{eq-approx}).
The similar error for $R_{K\eta^\prime}$ from Eq.~(\ref{eq-ratio4}) should be smaller since both the numerator and the denominator are calculated for the same resonance
within the same theoretical framework. Then, we give the estimation of the ratio $R_{K\eta^\prime}$ as $(20\pm8)\%$.
Still, it tells that the proportion for $K\eta^\prime$ pair in the $K^*_0(1430)$ decay is non-negligible, and it is not in conflict with
the presence of a dominant branching fraction for the $K\pi$ decay mode in view of the margin of existing errors in both experimental measurements and theoretical predictions.

There is no experimental information about $K_0^*(1430) \to K\eta$ except for the ratio $R^{exp}_{K\eta}=(9.2 \pm 2.5^{+1.0}_{-2.5})\%$
measured in the Dalitz plot analysis of $\eta_c\to K^+K^-\eta$ by the \textit{BABAR} collaboration~\cite{prd89-112004}.
This observation is not in complete agreement with the SU(3) expectation as mentioned in their own article.
According to the definitions in Eqs.~(\ref{eq-gKpi})$\sim$(\ref{eq-gKeta}) and the relevant parameters, it is easily to get
\begin{eqnarray} \label{Ratio-1430Keta}
 R_{K\eta}&=&\frac{\mathcal{B}(K^*_0(1430) \to K\eta)}{\mathcal{B}(K^*_0(1430) \to K\pi)}
 =\frac{g^2_{K_0^*K\eta}q^{K\eta}_0}{g^2_{K_0^*K\pi}q^{K\pi}_0} \nonumber\\
 &=&(\sqrt{\frac{1}{3}} \cos{\phi}-\sqrt{\frac{2}{3}} \sin{\phi})^2\frac{\sqrt{[m^2_{K_0^{*}}-(m_K+m_\eta)^2][m^2_{K_0^{*}}-(m_K-m_\eta)^2]}}{\sqrt{[m^2_{K_0^{*}}-(m_K+m_\pi)^2][m^2_{K_0^{*}}-(m_K-m_\pi)^2]}} \nonumber \\
 &=&0.39\%.
\end{eqnarray}
This value is much less than $\textit{BABAR}$'s result but agrees with the parametrization for the $K^*_0(1430)$ resonance in Refs.~\cite{plb632-471,
prd81-014002,prd78-052001,prd89-074030,prd104-072002,prd104-012016,JHEP08-196,plb653-1} where the $K\eta$ contribution is ignored and the coupling constant
$g_{K\eta}$ is set to be zero.
Moreover, this result is comparable to the data ${\mathcal B}(K^*_2(1430) \to K\eta)=(0.15^{+0.34}_{-0.10})\%$ by the PDG~\cite{PDG2022} which supports
the suppression of $K\eta$ branching fraction for the even-spin strange mesons under SU(3) with $\eta-\eta^{\prime}$ mixing.
It also indicates that the quasi-two-body $B$ decays with subprocess $K^*_0(1430) \to K \eta$ should have small branching ratios and are more difficult
to be observed experimentally.
Since the mass of $K^*_0(1430)$ is slightly smaller than $m_K+m_{\eta^\prime}$, it is not proper to calculate the ratio $R_{K\eta^\prime}$ by a formula
similar to Eq.~(\ref{Ratio-1430Keta}) directly.

\begin{figure}[tbp]
\centerline{\epsfxsize=8cm \epsffile{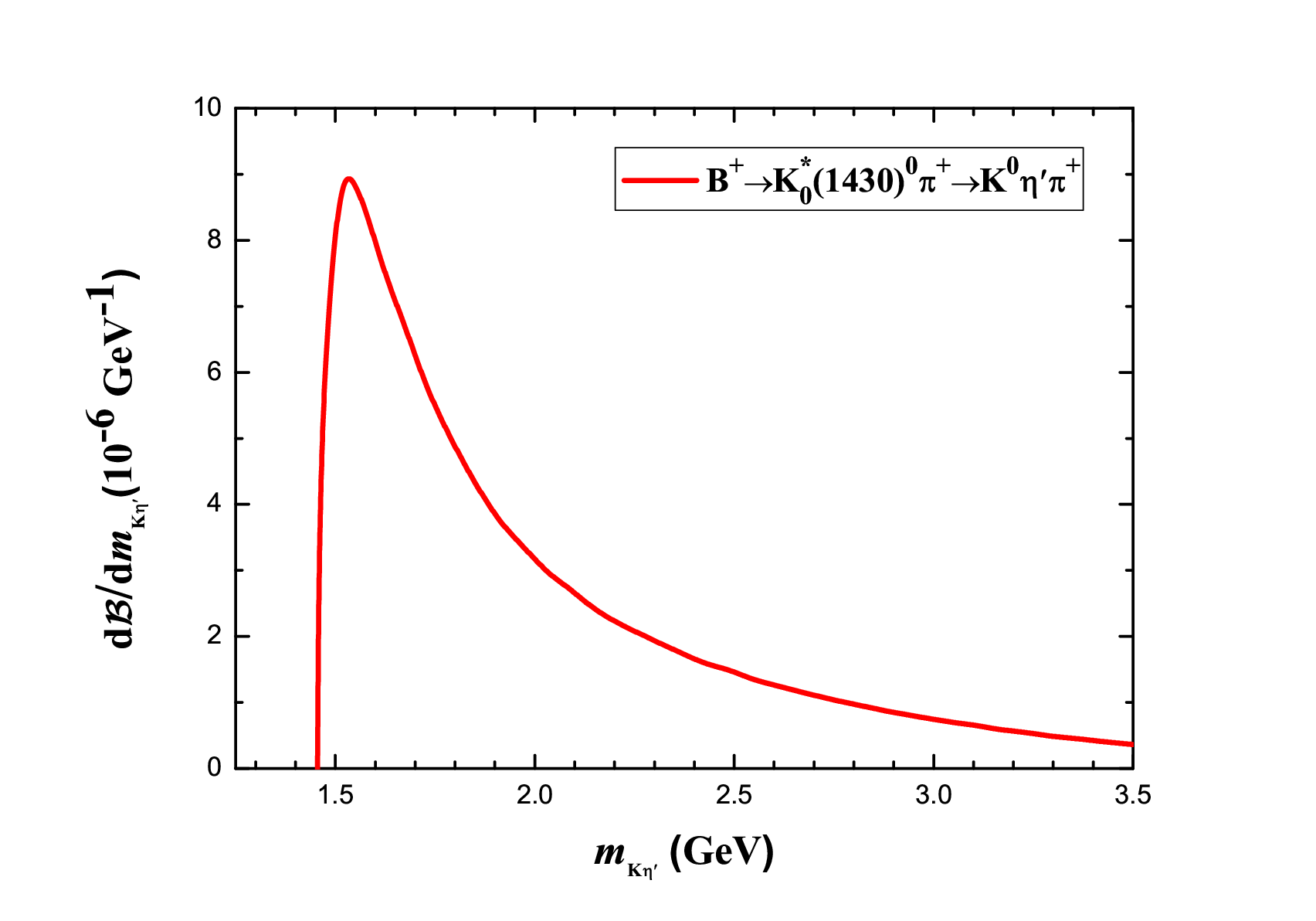}}
\caption{Differential branching fraction for the $B^+\to  K_0^{*}(1430)^0 \pi^+\to K^0\eta^{\prime} \pi^+$ decay with the invariant mass
$m_{K\eta^{\prime}}$ ranging from threshold to $3.5~\rm GeV$.}
\label{fig-diff}
\end{figure}

In Fig.~\ref{fig-diff}, we plot the differential branching fraction for the $B^+\to  K_0^{*}(1430)^0 \pi^+\to K^0\eta^{\prime} \pi^+$ decay with
the invariant mass $m_{K\eta^{\prime}}$ ranging from threshold to $3.5~\rm GeV$.
For a full resonant state contribution, the main portion of the branching fraction always lies in the region around the pole mass of the intermediate state.
But due to the strong suppression of phase space near threshold, one can find that the peak in the curve of the differential branching fraction for
the $B^+\to  K_0^{*}(1430)^0 \pi^+\to K^0\eta^{\prime} \pi^+$ decay is around $1.55~\rm GeV$ which cannot be considered as a new resonant contribution.
Similar behavior can also be seen in the $\eta^{\prime} K^\pm$ mass projection of the measured $\eta_c \to \eta^{\prime} K^+K^-$ Dalitz plot as
shown in Fig. 12(d) of Ref.~\cite{prd104-072002}.

%%%===========================%%%
\section{SUMMARY}
In this work, we studied the contributions from the subprocess $K^*_0(1430) \to K\eta^{\prime}$ in the charmless three-body $B\to K\eta^{\prime} h$ decays
by employing the PQCD approach.
In the description of the $K^*_0(1430)$ contribution, the related form factor was parametrized by the Flatt${\rm \acute{e}}$ formula with coupled channels
$K\pi$, $K\eta$, and $K\eta^{\prime}$, and the strong coupling constants $g_{K^*_0K\eta^{(\prime)}}$ were extracted from
$g_{K^*_0 K\pi}$ under the flavor SU$(3)$ symmetry.
The $CP$ averaged branching fractions for the $B\to K^*_0(1430) h \to K\eta^\prime h$ decays were predicted to be of the order of $10^{-8}$ to $10^{-5}$, which showed
the potential of experimental measurement.
In addition, the ratio between $\mathcal{B}(K^*_0(1430) \to K\eta^\prime)$ and $\mathcal{B}(K^*_0(1430) \to K\pi)$ was estimated to be about $20\%$
by the comparison of the branching fractions for corresponding quasi-two-body decays.
Since the coupling for $K^*_0(1430) \to K\eta$ is strongly suppressed within SU(3) symmetry, the quasi-two-body $B$ decays with subprocess $K^*_0(1430) \to K \eta$
are expected to have small branching ratios and have not been considered in this work.
There is no experimental measurement or theoretical analysis for the relevant three-body or quasi-two-body decays,
and the PQCD predictions are expected to be tested by the future experiments.

%-----------------------%
\begin{acknowledgments}
This work was supported by the National Natural Science Foundation of
China under Grants No. 12205148 and No. 11947011, the Natural Science Foundation of Jiangsu Province under
Grant No. BK20191010, the Qing Lan Project of Jiangsu Province, and the Fund for Shanxi ``1331 Project"
Key Subjects Construction.
\end{acknowledgments}
%-----------------------%


\begin{thebibliography}{199}
%\addtolength{\itemsep}{0.5ex}

%=============== Refs ===============%
%\cite{Close:2002zu}
\bibitem{jpg28-R249}
F.~E.~Close and N.~A.~T\"ornqvist,
%``Scalar mesons above and below 1-GeV,''
J. Phys. G \textbf{28}, R249 (2002).
%doi:10.1088/0954-3899/28/10/201
%[arXiv:hep-ph/0204205 [hep-ph]].
%440 citations counted in INSPIRE as of 15 May 2023


%\cite{Amsler:2004ps}
\bibitem{pr389-61}
C.~Amsler and N.~A.~T\"ornqvist,
%``Mesons beyond the naive quark model,''
Phys. Rep. \textbf{389}, 61 (2004).
%doi:10.1016/j.physrep.2003.09.003
%387 citations counted in INSPIRE as of 15 May 2023


%\cite{Bugg:2004xu}
\bibitem{pr397-257}
D.~V.~Bugg,
%``Four sorts of meson,''
Phys. Rep. \textbf{397}, 257 (2004).
%doi:10.1016/j.physrep.2004.03.008
%[arXiv:hep-ex/0412045 [hep-ex]].
%271 citations counted in INSPIRE as of 15 May 2023


%\cite{Klempt:2007cp}
\bibitem{pr454-1}
E.~Klempt and A.~Zaitsev,
%``Glueballs, Hybrids, Multiquarks. Experimental facts versus QCD inspired concepts,''
Phys. Rep. \textbf{454}, 1 (2007).
%doi:10.1016/j.physrep.2007.07.006
%[arXiv:0708.4016 [hep-ph]].
%814 citations counted in INSPIRE as of 15 May 2023


%\cite{Pelaez:2015qba}
\bibitem{pr658-1}
J.~R.~Pel\'aez,
%``From controversy to precision on the sigma meson: a review on the status of the non-ordinary $f_0(500)$ resonance,''
Phys. Rep. \textbf{658}, 1 (2016).
%doi:10.1016/j.physrep.2016.09.001
%[arXiv:1510.00653 [hep-ph]].
%357 citations counted in INSPIRE as of 15 May 2023


%\cite{Aston:1987ir}
\bibitem{npb296-493}
D.~Aston \textit{et al.} (LASS Collaboration), %N.~Awaji, T.~Bienz, F.~Bird, J.~D'Amore, W.~Dunwoodie, R.~Endorf, K.~Fujii, H.~Hayashi and S.~Iwata, \textit{et al.}
%``A Study of K- pi+ Scattering in the Reaction K- p ---\ensuremath{>} K- pi+ n at 11-GeV/c,''
Nucl. Phys. B\textbf{296}, 493 (1988).
%doi:10.1016/0550-3213(88)90028-4
%742 citations counted in INSPIRE as of 15 May 2023

%\cite{ParticleDataGroup:2022pth}
\bibitem{PDG2022}
R.~L.~Workman \textit{et al.} (Particle Data Group),
%``Review of Particle Physics,''
Prog. Theor. Exp. Phys. \textbf{2022}, 083C01 (2022).
%doi:10.1093/ptep/ptac097
%1001 citations counted in INSPIRE as of 15 May 2023

%\cite{BaBar:2014asx}
\bibitem{prd89-112004}
J.~P.~Lees \textit{et al.} (\textit{BABAR} Collaboration),
%``Dalitz plot analysis of $\eta_c \to K^+ K^- \eta$ and $\eta_c \to K^+ K^- \pi^0$ in two-photon interactions,''
Phys. Rev. D \textbf{89}, 112004 (2014).
%doi:10.1103/PhysRevD.89.112004
%[arXiv:1403.7051 [hep-ex]].
%41 citations counted in INSPIRE as of 17 May 2023


%\cite{BESIII:2014dlb}
\bibitem{prd89-074030}
M.~Ablikim \textit{et al.} (BESIII Collaboration),
%``Measurement of $\chi_{cJ}$ decaying into $\eta^{\prime}K^+K^-$,''
Phys. Rev. D \textbf{89}, 074030 (2014).
%doi:10.1103/PhysRevD.89.074030
%[arXiv:1402.2023 [hep-ex]].
%9 citations counted in INSPIRE as of 17 May 2023

%\cite{BaBar:2021fkz}
\bibitem{prd104-072002}
J.~P.~Lees \textit{et al.} (\textit{BABAR} Collaboration),
%``Light meson spectroscopy from Dalitz plot analyses of $\eta_c$ decays to $\eta' K^+ K^-$, $\eta' \pi^+ \pi^-$, and $\eta \pi^+ \pi^-$ produced in two-photon interactions,''
Phys. Rev. D \textbf{104}, 072002 (2021).
%doi:10.1103/PhysRevD.104.072002
%[arXiv:2106.05157 [hep-ex]].
%23 citations counted in INSPIRE as of 17 May 2023

%%%%%%%%%%%%%%%%%%belle
%\cite{Belle:2004drb}
\bibitem{prd71-092003}
A.~Garmash \textit{et al.} (Belle Collaboration),
%``Dalitz analysis of the three-body charmless decays B+ ---\ensuremath{>} K+ pi+ pi- and B+ ---\ensuremath{>} K+ K+ K-,''
Phys. Rev. D \textbf{71}, 092003 (2005).
%doi:10.1103/PhysRevD.71.092003
%[arXiv:hep-ex/0412066 [hep-ex]].
%221 citations counted in INSPIRE as of 22 May 2023

%\cite{Belle:2006ljg}
\bibitem{prd75-012006}
A.~Garmash \textit{et al.} (Belle Collaboration),
%``Dalitz Analysis of Three-body Charmless B0 ---\ensuremath{>} K0 pi+ pi- Decay,''
Phys. Rev. D \textbf{75}, 012006 (2007).
%doi:10.1103/PhysRevD.75.012006
%[arXiv:hep-ex/0610081 [hep-ex]].
%89 citations counted in INSPIRE as of 22 May 2023

%\cite{Belle:2005rpz}
\bibitem{prl96-251803}
A.~Garmash \textit{et al.} (Belle Collaboration),
%``Evidence for large direct CP violation in B+- ---\ensuremath{>} rho(770)0K+- from analysis of the three-body charmless B+- ---\ensuremath{>}K+- pi+- pi-+ decay,''
Phys. Rev. Lett. \textbf{96}, 251803 (2006).
%doi:10.1103/PhysRevLett.96.251803
%[arXiv:hep-ex/0512066 [hep-ex]].
%186 citations counted in INSPIRE as of 22 May 2023

%%%%%%%%%%%%%%%%%%BABAR
%\cite{BaBar:2005qms}
\bibitem{prd72-072003}
B.~Aubert \textit{et al.} (\textit{BABAR} Collaboration),
%``Dalitz-plot analysis of the decays $B^\pm \to K^\pm \pi^\mp \pi^\pm$,''
Phys. Rev. D \textbf{72}, 072003 (2005); \textbf{74}, 099903(E) (2006).
%doi:10.1103/PhysRevD.72.072003
%[arXiv:hep-ex/0507004 [hep-ex]].
%156 citations counted in INSPIRE as of 22 May 2023

%\cite{BaBar:2007eog}
\bibitem{prd76-071103}
B.~Aubert \textit{et al.} (\textit{BABAR} Collaboration),
%``Search for the decay $B^{+} \to \bar{K}$* 0(892) $K^{+}$,''
Phys. Rev. D \textbf{76}, 071103(R) (2007).
%doi:10.1103/PhysRevD.76.071103
%[arXiv:0706.1059 [hep-ex]].
%15 citations counted in INSPIRE as of 22 May 2023

%\cite{BaBar:2008lpx}
\bibitem{prd78-012004}
B.~Aubert \textit{et al.} (\textit{BABAR} Collaboration),
%``Evidence for Direct CP Violation from Dalitz-plot analysis of $B^\pm \to K^\pm \pi^\mp \pi^\pm$,''
Phys. Rev. D \textbf{78}, 012004 (2008).
%doi:10.1103/PhysRevD.78.012004
%[arXiv:0803.4451 [hep-ex]].
%144 citations counted in INSPIRE as of 22 May 2023

%\cite{BaBar:2009jov}
\bibitem{prd80-112001}
B.~Aubert \textit{et al.} (\textit{BABAR} Collaboration),
%``Time-dependent amplitude analysis of B0 ---\ensuremath{>} K0(S) pi+ pi-,''
Phys. Rev. D \textbf{80}, 112001 (2009).
%doi:10.1103/PhysRevD.80.112001
%[arXiv:0905.3615 [hep-ex]].
%97 citations counted in INSPIRE as of 22 May 2023

%\cite{BaBar:2011vfx}
\bibitem{prd83-112010}
J.~P.~Lees \textit{et al.} (\textit{BABAR} Collaboration),
%``Amplitude Analysis of $B^0\to K^+ \pi^- \pi^0$ and Evidence of Direct CP Violation in $B\to K^* \pi$ decays,''
Phys. Rev. D \textbf{83}, 112010 (2011).
%doi:10.1103/PhysRevD.83.112010
%[arXiv:1105.0125 [hep-ex]].
%52 citations counted in INSPIRE as of 22 May 2023

%\cite{BaBar:2015pwa}
\bibitem{prd96-072001}
J.~P.~Lees \textit{et al.} (\textit{BABAR} Collaboration),
%``Evidence for $CP$ violation in $B^{+} \to K^{*}(892)^{+} \pi^{0}$ from a Dalitz plot analysis of $B^{+} \to K^{0}_{S} \pi^{+} \pi^{0}$ decays,''
Phys. Rev. D \textbf{96}, 072001 (2017).
%doi:10.1103/PhysRevD.96.072001
%[arXiv:1501.00705 [hep-ex]].
%27 citations counted in INSPIRE as of 22 May 2023

%%%%%%%%%%%%%%%%%%LHCb
%\cite{LHCb:2019xmb}
\bibitem{prl123-231802}
R.~Aaij \textit{et al.} (LHCb Collaboration),
%``Amplitude analysis of $B^{\pm} \to \pi^{\pm} K^{+} K^{-}$ decays,''
Phys. Rev. Lett. \textbf{123}, 231802 (2019).
%doi:10.1103/PhysRevLett.123.231802
%[arXiv:1905.09244 [hep-ex]].
%56 citations counted in INSPIRE as of 22 May 2023

%\cite{LHCb:2019vww}
\bibitem{JHEP06-114}
R.~Aaij \textit{et al.} (LHCb Collaboration),
%``Amplitude analysis of $B^{0}_{s} \rightarrow K^{0}_{\textrm{S}} K^{\pm}\pi^{\mp}$ decays,''
J. High Energy Phys. 06 (2019) 114.
%doi:10.1007/JHEP06(2019)114
%[arXiv:1902.07955 [hep-ex]].
%21 citations counted in INSPIRE as of 22 May 2023

%%%%%theory
%\cite{Chen:2002th}
\bibitem{plb561-258}
C.~H.~Chen and H.~n.~Li,
%``Three body nonleptonic B decays in perturbative QCD,''
Phys. Lett. B \textbf{561}, 258 (2003).
%doi:10.1016/S0370-2693(03)00486-6
%[arXiv:hep-ph/0209043 [hep-ph]].
%95 citations counted in INSPIRE as of 27 May 2023

%\cite{Krankl:2015fha}
\bibitem{npb899-247}
S.~Kr\"ankl, T.~Mannel, and J.~Virto,
%``Three-body non-leptonic B decays and QCD factorization,''
Nucl. Phys. B\textbf{899}, 247 (2015).
%doi:10.1016/j.nuclphysb.2015.08.004
%[arXiv:1505.04111 [hep-ph]].
%87 citations counted in INSPIRE as of 27 May 2023

%\cite{Klein:2017xti}
\bibitem{JHEP10-117}
R.~Klein, T.~Mannel, J.~Virto, and K.~K.~Vos,
%``CP Violation in Multibody $B$ Decays from QCD Factorization,''
J. High Energy Phys. 10 (2017) 117.
%doi:10.1007/JHEP10(2017)117
%[arXiv:1708.02047 [hep-ph]].
%51 citations counted in INSPIRE as of 27 May 2023

%\cite{El-Bennich:2009gqk}
\bibitem{prd79-094005}
B.~El-Bennich, A.~Furman, R.~Kami\'nski, L.~Le\'sniak, B.~Loiseau, and B.~Moussallam,
%``CP violation and kaon-pion interactions in B ---\ensuremath{>} K pi+ pi- decays,''
Phys. Rev. D \textbf{79}, 094005 (2009); \textbf{83}, 039903(E) (2011).
%doi:10.1103/PhysRevD.83.039903
%[arXiv:0902.3645 [hep-ph]].
%109 citations counted in INSPIRE as of 26 May 2023

%\cite{Leitner:2010ai}
\bibitem{prd81-094033}
O.~Leitner, J.~P.~Dedonder, B.~Loiseau, and R.~Kami\'nski,
%``K* resonance effects on direct CP violation in B -\ensuremath{>} pi pi K,''
Phys. Rev. D \textbf{81}, 094033 (2010); \textbf{82}, 119906(E) (2010).
%doi:10.1103/PhysRevD.81.094033
%[arXiv:1001.5403 [hep-ph]].
%25 citations counted in INSPIRE as of 26 May 2023

%\cite{Cheng:2013dua}
\bibitem{prd88-114014}
H.~Y.~Cheng and C.~K.~Chua,
%``Branching Fractions and Direct CP Violation in Charmless Three-body Decays of B Mesons,''
Phys. Rev. D \textbf{88}, 114014 (2013).
%doi:10.1103/PhysRevD.88.114014
%[arXiv:1308.5139 [hep-ph]].
%88 citations counted in INSPIRE as of 26 May 2023

%\cite{Cheng:2014uga}
\bibitem{prd89-074025}
H.~Y.~Cheng and C.~K.~Chua,
%``Charmless three-body decays of $B_s$ mesons,''
Phys. Rev. D \textbf{89}, 074025 (2014).
%doi:10.1103/PhysRevD.89.074025
%[arXiv:1401.5514 [hep-ph]].
%42 citations counted in INSPIRE as of 26 May 2023

%\cite{Cheng:2016shb}
\bibitem{prd94-094015}
H.~Y.~Cheng, C.~K.~Chua, and Z.~Q.~Zhang,
%``Direct CP Violation in Charmless Three-body Decays of $B$ Mesons,''
Phys. Rev. D \textbf{94}, 094015 (2016).
%doi:10.1103/PhysRevD.94.094015
%[arXiv:1607.08313 [hep-ph]].
%65 citations counted in INSPIRE as of 26 May 2023

%\cite{Cheng:2020ipp}
\bibitem{prd102-053006}
H.~Y.~Cheng and C.~K.~Chua,
%``Branching fractions and $CP$ violation in $B^-\to K^+K^-\pi^-$ and $B^-\to \pi^+\pi^-\pi^-$ decays,''
Phys. Rev. D \textbf{102}, 053006 (2020).
%doi:10.1103/PhysRevD.102.053006
%[arXiv:2007.02558 [hep-ph]].
%17 citations counted in INSPIRE as of 26 May 2023

%\cite{Cheng:2020iwk}
\bibitem{prd103-036017}
H.~Y.~Cheng, C.~W.~Chiang, and C.~K.~Chua,
%``Finite-Width Effects in Three-Body B Decays,''
Phys. Rev. D \textbf{103}, 036017 (2021).
%doi:10.1103/PhysRevD.103.036017
%[arXiv:2011.07468 [hep-ph]].
%21 citations counted in INSPIRE as of 26 May 2023

%\cite{Li:2014oca}
\bibitem{prd89-094007}
Y.~Li,
%``Comprehensive study of $\overline B^0\to K^0(\overline K^0) K^\mp\pi^\pm$ decays in the factorization approach,''
Phys. Rev. D \textbf{89}, 094007 (2014).
%doi:10.1103/PhysRevD.89.094007
%[arXiv:1402.6052 [hep-ph]].
%42 citations counted in INSPIRE as of 26 May 2023

%\cite{Qi:2018syl}
\bibitem{prd99-076010}
J.~J.~Qi, X.~H.~Guo, Z.~Y.~Wang, Z.~H.~Zhang, and C.~Wang,
%``Study of $CP$ Violation in $B^-\rightarrow K^- \pi^+\pi^-$ and $B^-\rightarrow K^- \sigma(600)$ decays in the QCD factorization approach,''
Phys. Rev. D \textbf{99}, 076010 (2019).
%doi:10.1103/PhysRevD.99.076010
%[arXiv:1811.02167 [hep-ph]].
%17 citations counted in INSPIRE as of 26 May 2023

%\cite{Wang:2020saq}
\bibitem{JHEP03-162}
W.~F.~Wang, J.~Chai, and A.~J.~Ma,
%``Contributions of $K^*_0(1430)$ and $K^*_0(1950)$ in the three-body decays $B\to K\pi h$,''
J. High Energy Phys. 03 (2020) 162.
%doi:10.1007/JHEP03(2020)162
%[arXiv:2001.00355 [hep-ph]].
%16 citations counted in INSPIRE as of 26 May 2023

%\cite{Zou:2020fax}
\bibitem{epjc80-517}
Z.~T.~Zou, Y.~Li, and X.~Liu,
%``Branching fractions and CP asymmetries of the quasi-two-body decays in $B_{s} \rightarrow K^0({\overline{K}}^0)K^\pm \pi ^\mp $ within PQCD approach,''
Eur. Phys. J. C \textbf{80}, 517 (2020).
%doi:10.1140/epjc/s10052-020-8094-4
%[arXiv:2005.02097 [hep-ph]].
%16 citations counted in INSPIRE as of 26 May 2023

%\cite{Shi:2021ste}
\bibitem{epjc82-113}
Y.~J.~Shi, U.~G.~Mei\ss{}ner, and Z.~X.~Zhao,
%``Resonance contributions in $B^-\rightarrow K^+K^-\pi ^-$ within the light-cone sum rule approach,''
Eur. Phys. J. C \textbf{82}, 113 (2022).
%doi:10.1140/epjc/s10052-022-10062-0
%[arXiv:2111.05647 [hep-ph]].
%4 citations counted in INSPIRE as of 26 May 2023

%\cite{Lipkin:1980tk}
\bibitem{prl46-1307}
H.~J.~Lipkin,
%``The Importance of the $K_{\eta}$ and $K_{\eta}^\prime$ Decay Modes in Understanding Charmed and Other Meson Decays,''
Phys. Rev. Lett. \textbf{46}, 1307 (1981).
%doi:10.1103/PhysRevLett.46.1307
%62 citations counted in INSPIRE as of 21 Oct 2023

%\cite{Lipkin:1990us}
\bibitem{plb254-247}
H.~J.~Lipkin,
%``Interference effects in K eta and K eta-prime decay modes of heavy mesons: Clues to understanding weak transitions and CP violation,''
Phys. Lett. B \textbf{254}, 247 (1991).
%doi:10.1016/0370-2693(91)90429-T
%129 citations counted in INSPIRE as of 21 Oct 2023

%\cite{Aston:1987ey}
\bibitem{plb201-169}
D.~Aston \textit{et al.}(LASS Collaboration), %N.~Awaji, T.~Bienz, F.~Bird, J.~D'Amore, W.~Dunwoodie, R.~Endorf, K.~Fujii, H.~Hayashii and S.~Iwata, \textit{et al.}
%``Observation of the Selective Coupling of $K^*$ States to the $K^- \eta$ Channel,''
Phys. Lett. B \textbf{201}, 169 (1988).
%doi:10.1016/0370-2693(88)90102-5
%31 citations counted in INSPIRE as of 21 Oct 2023

%\cite{Bugg:2005xx}
\bibitem{plb632-471}
D.~V.~Bugg,
%``The Kappa in E791 data for D ---\ensuremath{>} K pi pi,''
Phys. Lett. B \textbf{632}, 471 (2006).
%doi:10.1016/j.physletb.2005.11.019
%[arXiv:hep-ex/0510019 [hep-ex]].
%58 citations counted in INSPIRE as of 05 Jun 2023

%\cite{CLEO:2008jus}
\bibitem{prd78-052001}
G.~Bonvicini \textit{et al.} (CLEO Collaboration),
%``Dalitz plot analysis of the D+ ---\ensuremath{>} K- pi+ pi+ decay,''
Phys. Rev. D \textbf{78}, 052001 (2008).
%doi:10.1103/PhysRevD.78.052001
%[arXiv:0802.4214 [hep-ex]].
%94 citations counted in INSPIRE as of 05 Jun 2023

%\cite{Bugg:2009uk}
\bibitem{prd81-014002}
D.~V.~Bugg,
%``An Update on the Kappa,''
Phys. Rev. D \textbf{81}, 014002 (2010).
%doi:10.1103/PhysRevD.81.014002
%[arXiv:0906.3992 [hep-ph]].
%33 citations counted in INSPIRE as of 10 Jun 2023

%\cite{BESIII:2020ctr}
\bibitem{prd104-012016}
M.~Ablikim \textit{et al.} (BESIII Collaboration),
%``Amplitude analysis and branching fraction measurement of $D_{s}^{+} \rightarrow K^{+}K^{-}\pi^{+}$,''
Phys. Rev. D \textbf{104}, 012016 (2021).
%doi:10.1103/PhysRevD.104.012016
%[arXiv:2011.08041 [hep-ex]].
%35 citations counted in INSPIRE as of 21 Oct 2023

%\cite{BESIII:2022vaf}
\bibitem{JHEP08-196}
M.~Ablikim \textit{et al.} (BESIII Collaboration),
%``Amplitude analysis and branching fraction measurement of the decay $D_{s}^{+} \to K^+\pi^+\pi^-$,''
J. High Energy Phys. 08 (2022) 196.
%doi:10.1007/JHEP08(2022)196
%[arXiv:2205.08844 [hep-ex]].
%5 citations counted in INSPIRE as of 21 Oct 2023

%\cite{FOCUS:2007mcb}
\bibitem{plb653-1}
J.~M.~Link \textit{et al.} (FOCUS Collaboration),
%``Dalitz plot analysis of the $D^{+} \to K^{-} \pi^{+} \pi^{+}$ decay in the FOCUS experiment,''
Phys. Lett. B \textbf{653}, 1 (2007).
%doi:10.1016/j.physletb.2007.06.070
%[arXiv:0705.2248 [hep-ex]].
%90 citations counted in INSPIRE as of 22 Oct 2023

%\cite{Keum:2000wi}
\bibitem{prd63-054008}
Y.~Y.~Keum, H.~n.~Li, and A.~I.~Sanda,
%``Penguin enhancement and $B \to K \pi$ decays in perturbative QCD,''
Phys. Rev. D \textbf{63}, 054008 (2001).
%doi:10.1103/PhysRevD.63.054008
%[arXiv:hep-ph/0004173 [hep-ph]].
%844 citations counted in INSPIRE as of 01 Jun 2023

%\cite{Keum:2000ph}
\bibitem{plb504-6}
Y.~Y.~Keum, H.~n.~Li, and A.~I.~Sanda,
%``Fat penguins and imaginary penguins in perturbative QCD,''
Phys. Lett. B \textbf{504}, 6 (2001).
%doi:10.1016/S0370-2693(01)00247-7
%[arXiv:hep-ph/0004004 [hep-ph]].
%737 citations counted in INSPIRE as of 01 Jun 2023

%\cite{Lu:2000em}
\bibitem{prd63-074009}
C.~D.~L\"u, K.~Ukai, and M.~Z.~Yang,
%``Branching ratio and CP violation of B ---\ensuremath{>} pi pi decays in perturbative QCD approach,''
Phys. Rev. D \textbf{63}, 074009 (2001).
%doi:10.1103/PhysRevD.63.074009
%[arXiv:hep-ph/0004213 [hep-ph]].
%552 citations counted in INSPIRE as of 01 Jun 2023

%%%%%%%%%%%%%PQCD-Q2B%%%%%
%\cite{Wang:2015uea}
\bibitem{prd91-042024}
W.~F.~Wang, H.~n.~Li, W.~Wang, and C.~D.~L\"u,
%``$S$-wave resonance contributions to the $B^0_{(s)}\to J/\psi\pi^+\pi^-$ and $B_s\to\pi^+\pi^-\mu^+\mu^-$ decays,''
Phys. Rev. D \textbf{91}, 094024 (2015).
%doi:10.1103/PhysRevD.91.094024
%[arXiv:1502.05483 [hep-ph]].
%81 citations counted in INSPIRE as of 01 Jun 2023

%\cite{Wang:2016rlo}
\bibitem{plb763-29}
W.~F.~Wang and H.~n.~Li,
%``Quasi-two-body decays $B\to K\rho\to K\pi\pi$ in perturbative QCD approach,''
Phys. Lett. B \textbf{763}, 29 (2016).
%doi:10.1016/j.physletb.2016.10.026
%[arXiv:1609.04614 [hep-ph]].
%62 citations counted in INSPIRE as of 01 Jun 2023

%\cite{Wang:2020plx}
\bibitem{prd101-111901}
W.~F.~Wang,
%``Will the subprocesses $\rho(770,1450)^0\to K^+K^-$ contribute large branching fractions for $B^\pm \to \pi^\pm K^+K^-$ decays?,''
Phys. Rev. D \textbf{101}, 111901(R) (2020).
%doi:10.1103/PhysRevD.101.111901
%[arXiv:2004.09027 [hep-ph]].
%15 citations counted in INSPIRE as of 01 Jun 2023

%\cite{Wang:2020nel}
\bibitem{prd103-056021}
W.~F.~Wang,
%``Contributions for the kaon pair from $\rho(770)$, $\omega(782)$ and their excited states in the $B\to K\bar K h$ decays,''
Phys. Rev. D \textbf{103}, 056021 (2021).
%doi:10.1103/PhysRevD.103.056021
%[arXiv:2012.15039 [hep-ph]].
%9 citations counted in INSPIRE as of 01 Jun 2023

%\cite{Fan:2020gvr}
\bibitem{epjc80-815}
Y.~Y.~Fan and W.~F.~Wang,
%``Resonance contributions $\phi (1020, 1680)\rightarrow K\bar{K}$ for the three-body decays $B\rightarrow K\bar{K} h$,''
Eur. Phys. J. C \textbf{80}, 815 (2020).
%doi:10.1140/epjc/s10052-020-8404-x
%[arXiv:2006.08223 [hep-ph]].
%7 citations counted in INSPIRE as of 01 Jun 2023

%\cite{Rui:2017bgg}
\bibitem{epjc77-199}
Z.~Rui, Y.~Li, and W.~F.~Wang,
%``The S-wave resonance contributions in the $B^0_s$ decays into $ \psi(2S,3S)$ plus pion pair,''
Eur. Phys. J. C \textbf{77}, 199 (2017).
%doi:10.1140/epjc/s10052-017-4772-2
%[arXiv:1701.02941 [hep-ph]].
%32 citations counted in INSPIRE as of 01 Jun 2023

%\cite{Rui:2017hks}
\bibitem{prd97-033006}
Z.~Rui and W.~F.~Wang,
%``$S$-wave $K\pi$ contributions to the hadronic charmonium $B$ decays in the perturbative QCD approach,''
Phys. Rev. D \textbf{97}, 033006 (2018).
%doi:10.1103/PhysRevD.97.033006
%[arXiv:1711.08959 [hep-ph]].
%24 citations counted in INSPIRE as of 01 Jun 2023

%\cite{Rui:2018hls}
\bibitem{prd98-113003}
Z.~Rui, Y.~Li, and H.~n.~Li,
%``$P$-wave contributions to $B\to\psi\pi\pi$ decays in perturbative QCD approach,''
Phys. Rev. D \textbf{98}, 113003 (2018).
%doi:10.1103/PhysRevD.98.113003
%[arXiv:1809.04754 [hep-ph]].
%22 citations counted in INSPIRE as of 01 Jun 2023

%\cite{Rui:2019yxx}
\bibitem{epjc79-792}
Z.~Rui, Y.~Li, and H.~Li,
%``Studies of the resonance components in the $B_s$ decays into charmonia plus kaon pair,''
Eur. Phys. J. C \textbf{79}, 792 (2019).
%doi:10.1140/epjc/s10052-019-7313-3
%[arXiv:1907.04128 [hep-ph]].
%15 citations counted in INSPIRE as of 01 Jun 2023

%\cite{Zou:2020atb}
\bibitem{epjc80-394}
Z.~T.~Zou, Y.~Li, Q.~X.~Li, and X.~Liu,
%``Resonant contributions to three-body $B\rightarrow KKK$ decays in perturbative QCD approach,''
Eur. Phys. J. C \textbf{80}, 394 (2020).
%doi:10.1140/epjc/s10052-020-7925-7
%[arXiv:2003.03754 [hep-ph]].
%24 citations counted in INSPIRE as of 01 Jun 2023

%\cite{Zou:2022xrr}
\bibitem{epjc82-1076}
Z.~T.~Zou, W.~S.~Fang, X.~Liu, and Y.~Li,
%``Analysis of CKM-favored quasi-two-body $B \rightarrow D (R\rightarrow ) K \pi $ decays in PQCD approach,''
Eur. Phys. J. C \textbf{82}, 1076 (2022).
%doi:10.1140/epjc/s10052-022-11060-y
%[arXiv:2210.08522 [hep-ph]].
%1 citations counted in INSPIRE as of 01 Jun 2023

%\cite{Li:2016tpn}
\bibitem{prd95-054008}
Y.~Li, A.~J.~Ma, W.~F.~Wang, and Z.~J.~Xiao,
%``Quasi-two-body decays $B_{(s)}\to P\rho\to P\pi\pi$ in perturbative QCD approach,''
Phys. Rev. D \textbf{95}, 056008 (2017).
%doi:10.1103/PhysRevD.95.056008
%[arXiv:1612.05934 [hep-ph]].
%58 citations counted in INSPIRE as of 01 Jun 2023

%\cite{Li:2018psm}
\bibitem{epjc79-37}
Y.~Li, W.~F.~Wang, A.~J.~Ma, and Z.~J.~Xiao,
%``Quasi-two-body decays $B_{(s)}\to K^*(892)h\to K\pi h$ in perturbative QCD approach,''
Eur. Phys. J. C \textbf{79}, 37 (2019).
%doi:10.1140/epjc/s10052-019-6544-7
%[arXiv:1809.09816 [hep-ph]].
%27 citations counted in INSPIRE as of 01 Jun 2023

%\cite{Li:2019hnt}
\bibitem{prd101-016015}
Y.~Li, D.~C.~Yan, Z.~Rui, and Z.~J.~Xiao,
%``$S$, $P$ and $D$-wave resonance contributions to $B_{(s)} \to \eta_c(1S,2S) K\pi$ decays in the perturbative QCD approach,''
Phys. Rev. D \textbf{101}, 016015 (2020).
%doi:10.1103/PhysRevD.101.016015
%[arXiv:1911.09348 [hep-ph]].
%13 citations counted in INSPIRE as of 01 Jun 2023

%\cite{Ma:2016csn}
\bibitem{npb923-54}
A.~J.~Ma, Y.~Li, W.~F.~Wang, and Z.~J.~Xiao,
%``The quasi-two-body decays $B_{(s)} \to (D_{(s)},\bar{D}_{(s)}) \rho \to (D_{(s)}, \bar{D}_{(s)})\pi \pi$ in the perturbative QCD factorization approach,''
Nucl. Phys. B\textbf{923}, 54 (2017).
%doi:10.1016/j.nuclphysb.2017.07.014
%[arXiv:1611.08786 [hep-ph]].
%31 citations counted in INSPIRE as of 01 Jun 2023

%\cite{Ma:2019qlm}
\bibitem{epjc79-539}
A.~J.~Ma, W.~F.~Wang, Y.~Li, and Z.~J.~Xiao,
%``Quasi-two-body decays $B \to D K^*(892) \to D K \pi$ in the perturbative QCD approach,''
Eur. Phys. J. C \textbf{79}, 539 (2019).
%doi:10.1140/epjc/s10052-019-7055-2
%[arXiv:1901.03956 [hep-ph]].
%14 citations counted in INSPIRE as of 01 Jun 2023

%\cite{Ma:2020jsb}
\bibitem{prd103-016002}
A.~J.~Ma and W.~F.~Wang,
%``Contributions of the kaon pair from $\rho(770)$ for the three-body decays $B \to DK\bar K$,''
Phys. Rev. D \textbf{103}, 016002 (2021).
%doi:10.1103/PhysRevD.103.016002
%[arXiv:2010.12906 [hep-ph]].
%8 citations counted in INSPIRE as of 01 Jun 2023

%\cite{Chai:2021pyp}
\bibitem{prd103-033003}
J.~Chai, S.~Cheng, and A.~J.~Ma,
%``Probing isovector scalar mesons in the charmless three-body B decays,''
Phys. Rev. D \textbf{105}, 033003 (2022).
%doi:10.1103/PhysRevD.105.033003
%[arXiv:2109.00664 [hep-ph]].
%3 citations counted in INSPIRE as of 01 Jun 2023

%\cite{Yang:2022jog}
\bibitem{prd107-013001}
H.~Yang and X.~Q.~Yu,
%``Investigating Bs three-body decays of scalar mesons in perturbative QCD approach,''
Phys. Rev. D \textbf{107}, 013001 (2023).
%doi:10.1103/PhysRevD.107.013001
%[arXiv:2207.13228 [hep-ph]].
%0 citations counted in INSPIRE as of 01 Jun 2023

%\cite{Jamin:2001zq}
\bibitem{npb622-279}
M.~Jamin, J.~A.~Oller, and A.~Pich,
%``Strangeness changing scalar form-factors,''
Nucl. Phys. B\textbf{622}, 279 (2002).
%doi:10.1016/S0550-3213(01)00605-8
%[arXiv:hep-ph/0110193 [hep-ph]].
%197 citations counted in INSPIRE as of 17 Jul 2022

%\cite{Gonzalez-Solis:2019lze}
\bibitem{prd101-034010}
S.~Gonz\`alez-Sol\'\i{}s, A.~Miranda, J.~Rend\'on, and P.~Roig,
%``Effective-field theory analysis of the $\tau^{-}\to K^{-}(\eta^{(\prime)},K^{0}) \nu_{\tau}$ decays,''
Phys. Rev. D \textbf{101}, 034010 (2020).
%doi:10.1103/PhysRevD.101.034010
%[arXiv:1911.08341 [hep-ph]].
%11 citations counted in INSPIRE as of 17 Jul 2022

%\cite{Boito:2009qd}
\bibitem{prd80-054007}
D.~R.~Boito and R.~Escribano,
%``Kpi form-factors and final state interactions in D+ ---\ensuremath{>} K- pi+ pi+ decays,''
Phys. Rev. D \textbf{80}, 054007 (2009).
%doi:10.1103/PhysRevD.80.054007
%[arXiv:0907.0189 [hep-ph]].
%33 citations counted in INSPIRE as of 03 Jun 2023

%\cite{Maltman:1999jn}
\bibitem{plb462-14}
K.~Maltman,
%``The a0(980), a0(1450) and K0*(1430) scalar decay constants and the isovector scalar spectrum,''
Phys. Lett. B \textbf{462}, 14 (1999).
%doi:10.1016/S0370-2693(99)00913-2
%[arXiv:hep-ph/9906267 [hep-ph]].
%71 citations counted in INSPIRE as of 04 Jun 2023

%\cite{Feldmann:1998sh}
\bibitem{plb449-339}
T.~Feldmann, P.~Kroll, and B.~Stech,
%``Mixing and decay constants of pseudoscalar mesons: The Sequel,''
Phys. Lett. B \textbf{449}, 339 (1999).
%doi:10.1016/S0370-2693(99)00085-4
%[arXiv:hep-ph/9812269 [hep-ph]].
%363 citations counted in INSPIRE as of 05 Jun 2023

%\cite{Feldmann:1998vh}
\bibitem{prd58-114006}
T.~Feldmann, P.~Kroll, and B.~Stech,
%``Mixing and decay constants of pseudoscalar mesons,''
Phys. Rev. D \textbf{58}, 114006 (1998).
%doi:10.1103/PhysRevD.58.114006
%[arXiv:hep-ph/9802409 [hep-ph]].
%712 citations counted in INSPIRE as of 05 Jun 2023


%\cite{Flatte:1976xu}
\bibitem{plb63-224}
S.~M.~Flatt${\rm \acute{e}}$,
%``Coupled - Channel Analysis of the pi eta and K anti-K Systems Near K anti-K Threshold,''
Phys. Lett. \textbf{63}B, 224 (1976).
%doi:10.1016/0370-2693(76)90654-7
%637 citations counted in INSPIRE as of 05 Jun 2023



%\cite{Volkov:2022ukj}
\bibitem{epja59-79}
M.~K.~Volkov, K.~Nurlan, and A.~A.~Pivovarov,
%``On the decay widths of radially excited scalar meson $K^*_0(1430)$ in view of new experimental data,''
Eur. Phys. J. A \textbf{59}, 79 (2023).
%doi:10.1140/epja/s10050-023-00996-8
%[arXiv:2210.10557 [hep-ph]].
%1 citations counted in INSPIRE as of 10 Jun 2023

%\cite{Cheng:2005nb}
\bibitem{prd73-014017}
H.~Y.~Cheng, C.~K.~Chua, and K.~C.~Yang,
%``Charmless hadronic B decays involving scalar mesons: Implications to the nature of light scalar mesons,''
Phys. Rev. D \textbf{73}, 014017 (2006).
%doi:10.1103/PhysRevD.73.014017
%[arXiv:hep-ph/0508104 [hep-ph]].
%213 citations counted in INSPIRE as of 10 Jun 2023

%\cite{Meissner:2013hya}
\bibitem{plb730-336}
U.~G.~Mei\ss{}ner and W.~Wang,
%``Generalized Heavy-to-Light Form Factors in Light-Cone Sum Rules,''
Phys. Lett. B \textbf{730}, 336 (2014).
%doi:10.1016/j.physletb.2014.02.009
%[arXiv:1312.3087 [hep-ph]].
%67 citations counted in INSPIRE as of 10 Jun 2023

%\cite{Cheng:2013fba}
\bibitem{prd87-114001}
H.~Y.~Cheng, C.~K.~Chua, K.~C.~Yang, and Z.~Q.~Zhang,
%``Revisiting charmless hadronic B decays to scalar mesons,''
Phys. Rev. D \textbf{87}, 114001 (2013).
%doi:10.1103/PhysRevD.87.114001
%[arXiv:1303.4403 [hep-ph]].
%54 citations counted in INSPIRE as of 10 Jun 2023

%\cite{Ali:2007ff}
\bibitem{prd76-074018}
A.~Ali, G.~Kramer, Y.~Li, C.~D.~L\"u, Y.~L.~Shen, W.~Wang, and Y.~M.~Wang,
%``Charmless non-leptonic $B_s$ decays to $PP$, $PV$ and $VV$ final states in the pQCD approach,''
Phys. Rev. D \textbf{76}, 074018 (2007).
%doi:10.1103/PhysRevD.76.074018
%[arXiv:hep-ph/0703162 [hep-ph]].
%314 citations counted in INSPIRE as of 19 Jun 2023

%\cite{Bramon:1997va}
\bibitem{epjc7-271}
A.~Bramon, R.~Escribano, and M.~D.~Scadron,
%``The eta - eta-prime mixing angle revisited,''
Eur. Phys. J. C \textbf{7}, 271 (1999).
%doi:10.1007/s100529801009
%[arXiv:hep-ph/9711229 [hep-ph]].
%206 citations counted in INSPIRE as of 25 Jan 2024

%\cite{Cao:1999fs}
\bibitem{prd60-114012}
F.~G.~Cao and A.~I.~Signal,
%``Two analytical constraints on the eta - eta-prime mixing,''
Phys. Rev. D \textbf{60}, 114012 (1999).
%doi:10.1103/PhysRevD.60.114012
%[arXiv:hep-ph/9908481 [hep-ph]].
%33 citations counted in INSPIRE as of 25 Jan 2024

%\cite{Bramon:2000fr}
\bibitem{plb503-271}
A.~Bramon, R.~Escribano, and M.~D.~Scadron,
%``Radiative V P gamma transitions and eta - eta-prime mixing,''
Phys. Lett. B \textbf{503}, 271 (2001).
%doi:10.1016/S0370-2693(01)00161-7
%[arXiv:hep-ph/0012049 [hep-ph]].
%79 citations counted in INSPIRE as of 25 Jan 2024

%\cite{Xiao:2005af}
\bibitem{prd71-014034}
B.~W.~Xiao and B.~Q.~Ma,
%``Photon-meson transition form-factors of light pseudoscalar mesons,''
Phys. Rev. D \textbf{71}, 014034 (2005).
%doi:10.1103/PhysRevD.71.014034
%[arXiv:hep-ph/0501160 [hep-ph]].
%18 citations counted in INSPIRE as of 25 Jan 2024

%\cite{Escribano:2005qq}
\bibitem{jhep06-029}
R.~Escribano and J.~M.~Fr\`ere,
%``Study of the eta - eta-prime system in the two mixing angle scheme,''
J. High Energy Phys. 06 (2005) 029.
%doi:10.1088/1126-6708/2005/06/029
%[arXiv:hep-ph/0501072 [hep-ph]].
%170 citations counted in INSPIRE as of 25 Jan 2024

%\cite{Huang:2006as}
\bibitem{epjc50-771}
T.~Huang and X.~G.~Wu,
%``Determination of the eta and eta' Mixing Angle from the Pseudoscalar Transition Form Factors,''
Eur. Phys. J. C \textbf{50}, 771 (2007).
%doi:10.1140/epjc/s10052-007-0245-3
%[arXiv:hep-ph/0612007 [hep-ph]].
%24 citations counted in INSPIRE as of 25 Jan 2024

%\cite{Escribano:2007cd}
\bibitem{jhep05-006}
R.~Escribano and J.~Nadal,
%``On the gluon content of the eta and eta-prime mesons,''
J. High Energy Phys. 05 (2007) 006.
%doi:10.1088/1126-6708/2007/05/006
%[arXiv:hep-ph/0703187 [hep-ph]].
%109 citations counted in INSPIRE as of 25 Jan 2024

%\cite{KLOE:2006guu}
\bibitem{plb648-267}
F.~Ambrosino \textit{et al.} (KLOE Collaboration),
%``Measurement of the pseudoscalar mixing angle and eta-prime gluonium content with KLOE detector,''
Phys. Lett. B \textbf{648}, 267 (2007).
%doi:10.1016/j.physletb.2007.03.032
%[arXiv:hep-ex/0612029 [hep-ex]].
%141 citations counted in INSPIRE as of 25 Jan 2024

%\cite{Ambrosino:2009sc}
\bibitem{jhep07-105}
F.~Ambrosino \textit{et al.} (KLOE Collaboration),
%``A Global fit to determine the pseudoscalar mixing angle and the gluonium content of the eta-prime meson,''
J. High Energy Phys. 07 (2009) 105.
%doi:10.1088/1126-6708/2009/07/105
%[arXiv:0906.3819 [hep-ph]].
%152 citations counted in INSPIRE as of 25 Jan 2024

%\cite{LHCb:2014oms}
\bibitem{jhep01-024}
R.~Aaij \textit{et al.} (LHCb Collaboration),
%``Study of $\eta-\eta^{\prime}$ mixing from measurement of $B^0_{(s)} \rightarrow J/\psi \eta^{(\prime)}$\textasciitilde{}decay rates,''
J. High Energy Phys. 01 (2015) 024.
%doi:10.1007/JHEP01(2015)024
%[arXiv:1411.0943 [hep-ex]].
%37 citations counted in INSPIRE as of 25 Jan 2024

%\cite{BESIII:2023ajr}
\bibitem{prd108-092003}
M.~Ablikim \textit{et al.} (BESIII Collaboration),
%``Precision measurements of Ds+\textrightarrow{}\ensuremath{\eta}e+\ensuremath{\nu}e and Ds+\textrightarrow{}\ensuremath{\eta}'e+\ensuremath{\nu}e,''
Phys. Rev. D \textbf{108}, 092003 (2023).
%doi:10.1103/PhysRevD.108.092003
%[arXiv:2306.05194 [hep-ex]].
%3 citations counted in INSPIRE as of 25 Jan 2024

%\cite{BESIII:2023gbn}
\bibitem{2307.12852}
M.~Ablikim \textit{et al.} (BESIII Collaboration),
%``Observation of $D^+_s\to \eta^\prime \mu^+\nu_\mu$ and Measurements of $D^+_s\to \eta^{(\prime)}\mu^+\nu_\mu$ Decay Dynamics,''
arXiv:2307.12852.
%2 citations counted in INSPIRE as of 25 Jan 2024

%\cite{Liu:2010zg}
\bibitem{CTP53-540}
X.~Liu and Z.~J.~Xiao,
%``B ---\ensuremath{>} K*0(1430) K decays in perturbative QCD approach,''
Commun. Theor. Phys. \textbf{53}, 540 (2010).
%doi:10.1088/0253-6102/53/3/26
%[arXiv:1004.0749 [hep-ph]].
%16 citations counted in INSPIRE as of 17 Nov 2023

%\cite{Li:2015zra}
\bibitem{prd91-074022}
Y.~Li, H.~Y.~Zhang, Y.~Xing, Z.~H.~Li, and C.~D.~L\"u,
%``Study of $B \to K_0^*(1430)K^{(*)}$ decays in QCD Factorization Approach,''
Phys. Rev. D \textbf{91}, 074022 (2015).
%doi:10.1103/PhysRevD.91.074022
%[arXiv:1501.03865 [hep-ph]].
%8 citations counted in INSPIRE as of 11 Nov 2023

%\cite{Chen:2021oul}
\bibitem{prd105-016002}
L.~Chen, M.~Zhao, Y.~Zhang, and Q.~Chang,
%``Study of $B_{u,d,s} \to K^*_0$ (1430)$P$ and $K^*_0 (1430)V$ decays within QCD factorization,''
Phys. Rev. D \textbf{105}, 016002 (2022).
%doi:10.1103/PhysRevD.105.016002
%[arXiv:2112.00915 [hep-ph]].
%5 citations counted in INSPIRE as of 11 Nov 2023

%\cite{Cheng:2020mna}
\bibitem{plb813-136058}
H.~Y.~Cheng, C.~W.~Chiang, and C.~K.~Chua,
%``Width effects in resonant three-body decays: $B$ decay as an example,''
Phys. Lett. B \textbf{813}, 136058 (2021).
%doi:10.1016/j.physletb.2020.136058
%[arXiv:2011.03201 [hep-ph]].
%17 citations counted in INSPIRE as of 22 Jan 2024
%%%% ---ZZZZZZZ----------
\end{thebibliography}
\end{document}